\documentclass[twocolumn]{aastex63}
\makeatletter
\providecommand{\@LN}[2]{}
\makeatother
\usepackage{graphicx}
\usepackage{float}
\usepackage{nicefrac}
\usepackage{savesym}
\savesymbol{tablenum}
\usepackage{siunitx}
\restoresymbol{SIX}{tablenum}
\usepackage{appendix}
\usepackage{dcolumn}
\usepackage{bm}
\usepackage[displaymath]{lineno}
\usepackage{color,soul}
\usepackage{amsmath}
\usepackage{etoolbox} 

\newcommand*\linenomathpatch[1]{%
  \cspreto{#1}{\linenomath}%
  \cspreto{#1*}{\linenomath}%
  \cspreto{end#1}{\endlinenomath}%
  \cspreto{end#1*}{\endlinenomath}%
}

\linenomathpatch{equation}
\linenomathpatch{gather}
\linenomathpatch{multline}
\linenomathpatch{align}
\linenomathpatch{alignat}
\linenomathpatch{flalign}

\shorttitle{Radiation Effects for Interstellar Spacecraft}
\shortauthors{Lubin, Cohen, Erlikhman}
\graphicspath{{./}{figures/}}

\begin{document}

\title{Radiation Effects from ISM and Cosmic Ray Particle Impacts on Relativistic Spacecraft}

\author{Philip Lubin}
\affiliation{Department of Physics, University of California - Santa Barbara, Santa Barbara, CA, 93106, USA}

\author{Alexander N. Cohen}
\affiliation{Department of Physics, University of California - Santa Barbara, Santa Barbara, CA, 93106, USA}

\author{Jacob Erlikhman}
\affiliation{Department of Physics, University of California - Berkeley, Berkeley, CA, 94720, USA}



\begin{abstract}

Relativistic spacecraft, like those proposed by the NASA Starlight program and the Breakthrough Starshot Initiative, will have to survive radiation production that is unique when compared to that experienced by conventional spacecraft. In a relativistic interstellar spacecraft's reference frame, the interstellar medium (ISM) will look like a nearly mono-energetic beam of charged particles which impinges upon the leading edge of the spacecraft. Upon impact, ISM protons and electrons will travel characteristic lengths through the spacecraft shield and come to a stop via electronic and nuclear stopping mechanisms. As a result, bremsstrahlung photons will be produced within the spacecraft shield. In this work, we discuss the interstellar environment and its implications for radiation damage on relativistic spacecraft. We also explore expected radiation doses in terms of on-board device radiation tolerance.

\end{abstract}



\section*{\label{sec:intro}Introduction}

    We take a 1-10 km aperture, 100 GW class laser phased array as our baseline directed energy (DE) propulsion system \citep{lubin_roadmap_2016}. The DE propulsion system, among numerous other uses, can theoretically propel a 1 g spacecraft to $\sim0.25c$ or a 1 kg spacecraft to $\sim0.04c$. For a mission to the nearest star system, $\alpha$  Centauri, this technology could enable flight times of approximately 20 years for very small spacecraft. While this paper is general in its application to relativistic spacecraft, we focus on spacecraft geometries that are thin disks.
    
    As discussed by \citep{hoang_interaction_2017}, \citep{drobny_survivability_2020}, and \citep{drobny_damage_2021}, long-term damaging effects will be caused by spacecraft erosion and ISM gas implantation, both of which can lead to macroscopic morphological changes to the spacecraft structure and thus threaten the performance of any such mission. However, the production of radiation upon impact of charged ISM and cosmic ray particles also poses a serious threat to spacecraft survivability, electronic performance, and operation of experiments, especially for missions with biological payloads, and will be the subject of discussion in this study.
    
    We begin with a discussion of the interstellar medium (ISM) in Section \ref{section:interstellarmedium}, and show how it is largely isotropic in the ISM rest frame, but strongly peaked in the forward direction in the frame of a relativistic spacecraft. This suggests that for a spacecraft which has the capability to retain its attitude, a raised edge shield would be an effective method to mitigate ISM particle bombardment. In Section \ref{section:pendepths}, we discuss the penetration depths for electrons and protons (see Figure \ref{fig:dEdx_range_vs_energy}, which is a critical factor in estimating the production of secondary particles and therefore radiation doses inside the spacecraft. In Section \ref{section:brem}, we discuss the production of bremsstrahlung photons by electrons, protons, and dust grains. Section \ref{section:particleproduction} presents a general method for the calculation of secondary particles of various types for incoming ISM species such as electrons, protons, and helium. In Section \ref{section:cosmicrays}, we discuss cosmic rays and show how the energy deposition in the spacecraft is dominated by proton impacts (see Figure \ref{fig:GalacticCosmicRayComposition}) around 1 GeV. Finally, Section \ref{section:radiationdoses} contains a discussion on device radiation tolerances and we show that we can expect approximately $20\beta$ proton hits per year per exposed square {\AA} cell on the spacecraft or its reflector.

\section{Interstellar Medium}
\label{section:interstellarmedium}

    For a relativistic spacecraft, impacts with the ISM can lead to extremely large radiation doses. This needs to be considered in details to assure spacecraft survival. The ISM is made up of approximately 90\% protons by number with about 8\% Helium and other elements, such as C, N, O, Ne and Fe making up 10$^{-3}$ to 10$^{-4}$ fractionally by number. For comparison, the proton density and speed near the Earth from the solar wind (not including magnetic field trapping) are about 9 protons cm$^{-3}$ and 500 km s$^{-1}$, while for the ISM it is about 0.2 protons cm$^{-3}$ (HI + HII) with an average speed of 
    \begin{equation}
        v_{av} = \sqrt{8kT/\pi m}\sim13\hspace{1mm}\textrm{km s$^{-1}$},
    \end{equation}
    corresponding to $\beta\sim4.3\times10^{-5}$ for protons and 540 km s$^{-1}$ ($\beta\sim1.8\times10^{-3}$) for electrons, with an equivalent temperature of about 7500 K \citep{draine_physics_2011}.  The ISM elemental distribution is similar to the solar system distribution.
    
\subsection{ISM Boosting and Transverse Bombardment}
    
    While the average kinetic energy of the ISM protons and electrons is highly non-relativistic, in the ISM ``rest frame,'' the spacecraft of the type we are considering boosts the collision speeds to essentially the speed of the spacecraft. For example, at a spacecraft speed $\beta$, the kinetic energy of the colliding ISM particle becomes 
    \begin{equation}
        \textrm{KE}=\mathop{m_{0p}}c^2(\gamma-1)\sim\frac{1}{2}\mathop{\beta^2}\mathop{m_{0p}}c^2
    \end{equation}
    for modest $\beta$ (typically $<0.5$), where $\mathop{m_{0p}}c^2$ is the rest mass of the colliding particle. For a proton, the rest mass is 938 MeV, while for an electron it is 511 KeV. For a spacecraft at $\beta=0.2$, a colliding ISM proton is the equivalent of a $\sim19$ MeV proton and the electron is $\sim10$ KeV. A serious issue is the bombardment rate of the spacecraft. With an ISM particle density $n_p$ and a particle speed relative to the local stars of $v_p$, the bombardment flux (particle collisions per second per unit area when spacecraft is at rest) for monoenergetic particles of speed
    $v_p$ is
    \begin{equation}
        \Gamma_p= n_p v_p/4,
    \end{equation}
    and 
    \begin{equation}
        \Gamma_p = n_p v_{av}/4
    \end{equation}
    for a distribution of particles with probability distribution function $f(v)$. This can be seen by calculating the particle flux for a thermal distribution impinging on a surface at rest:
    
    \begin{align}
    \begin{split}
        \Gamma_p=&\int_v \iint_{2\pi}vf(v)\frac{n_p}{2}\cos(\theta)\frac{\mathop{d\Omega}}{2\pi}\mathop{dv}\\
        =&\frac{n_p}{2}\int_v vf(v)\mathop{dv}\iint_{2\pi}\cos(\theta)\frac{d\Omega}{2\pi}\\
        =&n_p  v_{av} \iint_{2\pi}\cos(\theta)\frac{\mathop{d\Omega}}{4\pi}\\
        =&\frac{n_p v_{av}}{4\pi}\iint_{2\pi}\cos(\theta)\sin(\theta)\mathop{d\theta} \mathop{d\phi}\\
        =&\frac{n_p v_{av}}{2}\int_0^{\frac{\pi}{2}}\cos(\theta)\sin(\theta)\mathop{d\theta}=\frac{1}{4}n_p v_{av},
    \end{split}
    \end{align}
    where
    \begin{equation}
        v_{av}\equiv\int_v vf(v)\mathop{dv}
    \end{equation}
    is the mean thermal speed and $f(v)$ is the normalized probability distribution function. For a thermal Maxwell-Boltzmann distribution,
    \begin{equation}
        v_{av}=\int_v vf(v)\mathop{dv}=\sqrt{8kT/\pi m}
    \end{equation}
    is the average thermal speed of a thermalized particle of mass $m$.

\subsection{ISM Particle Density Distribution Function}
    
    If the average particle density is $n_p$, we can write the density speed distribution function as $n(v)=n_p f(v)$. Since $f(v)$ is normalized with
    \begin{equation}
        \int_0^\infty f(v)\mathop{dv}=1,
    \end{equation}
    we can show that $n_p(v)$ is normalized by
    \begin{equation}
        \int_0^\infty n_p(v)\mathop{dv}=\int_0^\infty n_p f(v) \mathop{dv}=n_p.
    \end{equation}
    
\subsection{Transforming to Spacecraft Coordinate System}
    
\begin{figure}[h]
    \centering
    \begin{tabular}{c c}
         \includegraphics[width=0.2\textwidth]{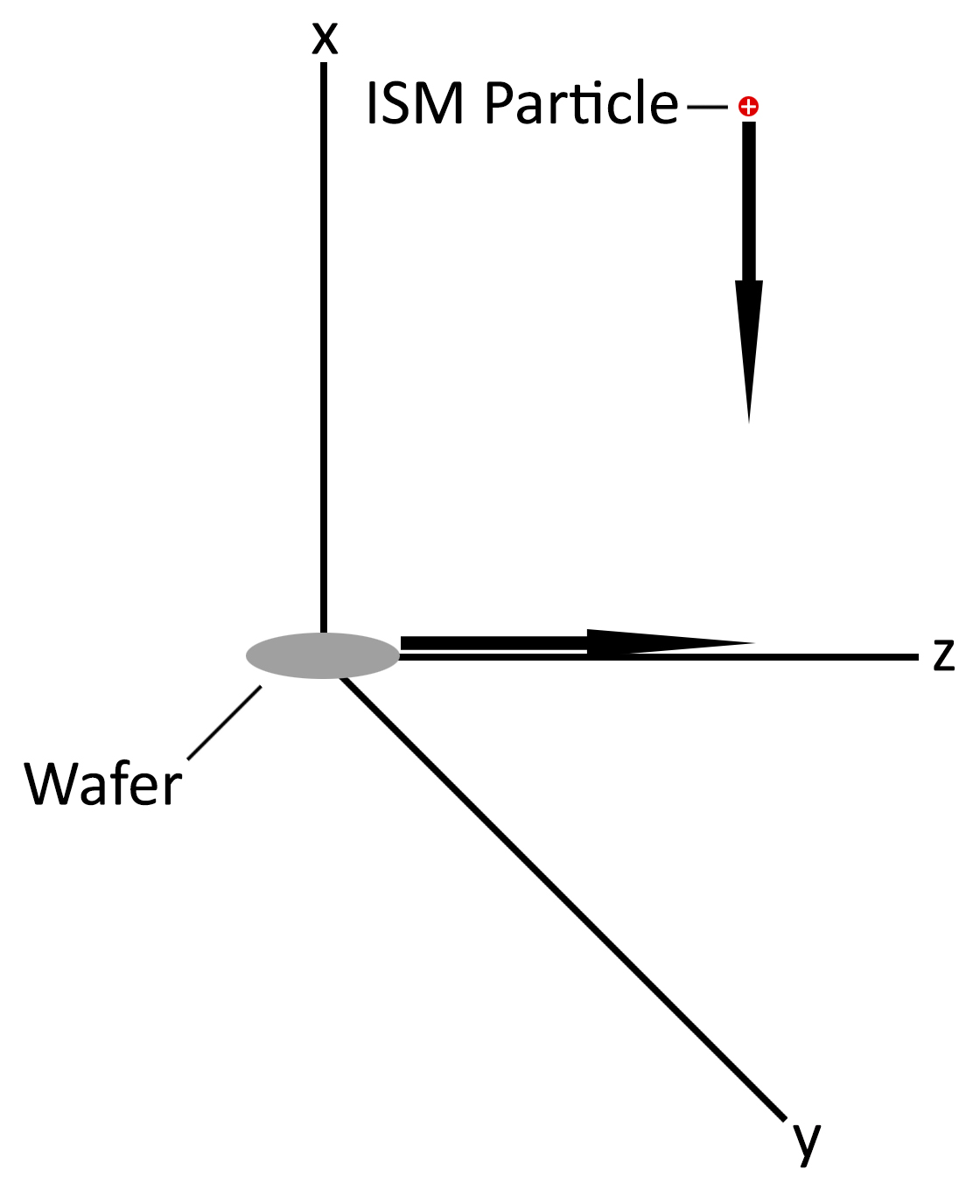} & \includegraphics[width=0.2\textwidth]{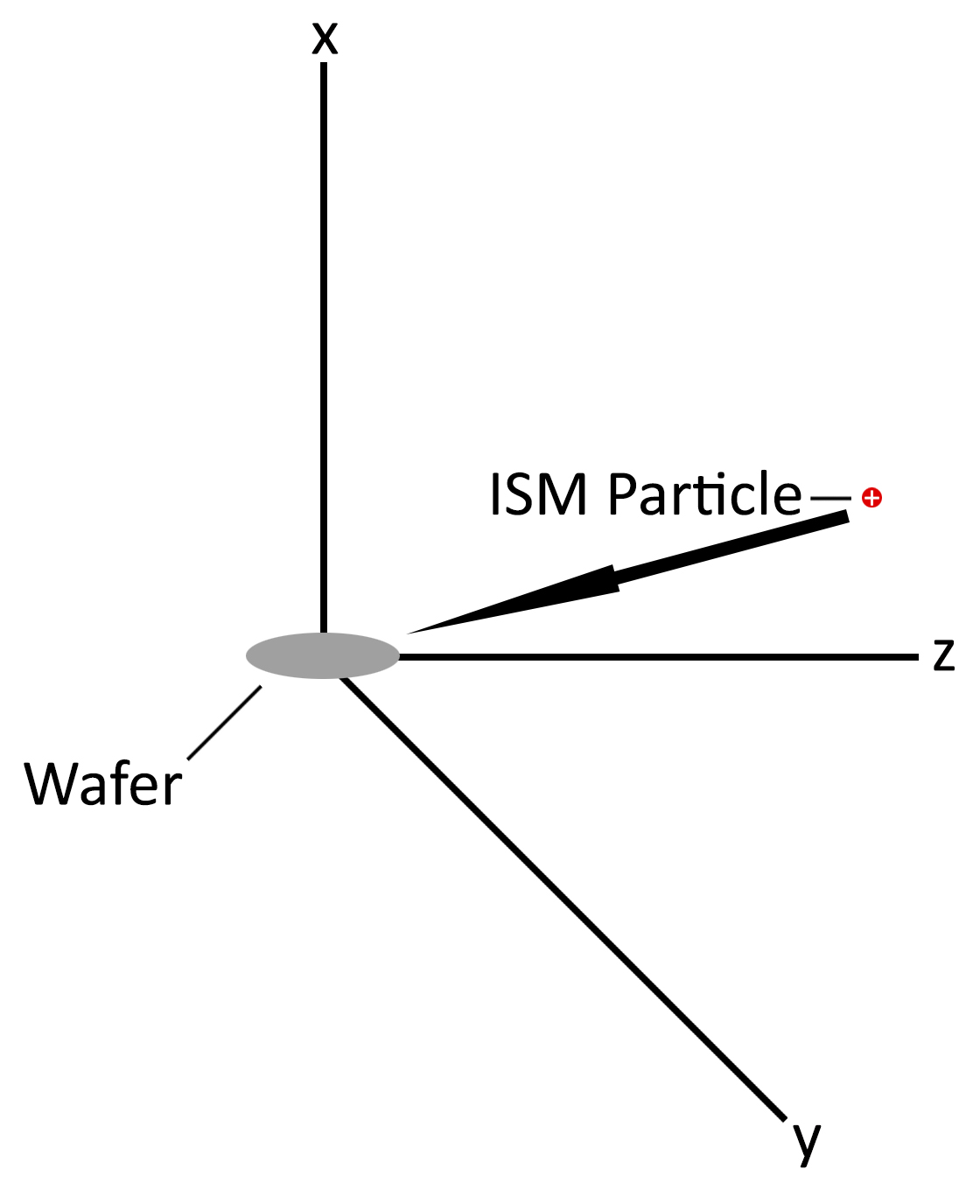}  \\
         (a) & (b) 
    \end{tabular}{}
    \caption{(a) Stationary ISM reference frame, with spacecraft shown as a thin grey wafer with the $x$-axis along it's axis of symmetry and a velocity vector along the $z$-axis. An example ISM particle is shown with arbitrary velocity vector. (b) Spacecraft reference frame (primed frame), wherein the composite ISM particle velocity vector is shown. As will be shown in this section, the ISM particle distribution function becomes strongly peaked in the forward direction in the reference frame of a relativistic spacecraft.}
    \label{fig:ISM_ref_frame2}
    \end{figure}

    The problem, of course, is the spacecraft is not at rest relative to the gas reference frame. The two reference frames of interest are illustrated in Figure \ref{fig:ISM_ref_frame2}.
    
    Assuming a mildly relativistic scenario, in the (primed) reference frame of the spacecraft we have a particle velocity given by 
    \begin{equation}
        \bm{v'}=(v'_x,v'_y,v'_z)=c\bm{\beta}-\bm{v},
    \end{equation}
    where $c\bm{\beta}$ is the spacecraft velocity in the ISM frame and $\bm{v}=(v_x,v_y,v_z)$ is the ISM frame particle velocity. Assuming spacecraft motion along the $z$-axis with 
    \begin{equation}
        c\bm{\beta}=(0,0,c\beta),
    \end{equation}
    the particle velocity in the spacecraft frame is then
    \begin{equation}
        \bm{v'}=c\bm{\beta}-\bm{v}=(v_x,v_y,c\beta-v_z).
    \end{equation}
    In the ISM frame, the particle velocity is then
    \begin{equation}
        \bm{v}=c\bm{\beta}-\bm{v'}=(v'_x,v'_y,c\beta-v'_z).
    \end{equation}
    
    \begin{widetext}
    For a thermalized ISM, the velocity vector distribution function for particles hitting the spacecraft is:
    \begin{align}
    \begin{split}
        f_v(v_x,v_y,v_z)=\bigg[\frac{m}{2\pi kT}\bigg]^{3/2}e^{-m(v_x^2+v_y^2+v_z^2)/2kT}=\bigg[\frac{m}{2\pi kT}\bigg]^{3/2}e^{-m\big[{v'_x}^2+{v'_y}^2+(c\beta-v'_z)^2\big]/2kT},
    \end{split}
    \end{align}
    with the speed distribution function being:
    
    \begin{align}
    \begin{split}
        f(v)\mathop{dv}=f_v(v_x,v_y,v_z)4\pi v^2\mathop{dv}=4\pi v^2 dv \bigg[\frac{m}{2\pi kT}\bigg]^{3/2}e^{-m\big[{v'_x}^2+{v'_y}^2+(c\beta-v'_z)^2\big]/2kT}.
    \end{split}
    \end{align}
    Therefore, we have
    \begin{align}
    \begin{split}
        f(v)=4\pi\big[{v'_x}^2+{v'_y}^2+(c\beta-v'_z)^2\big]\bigg[\frac{m}{2\pi kT}\bigg]^{3/2}e^{-m\big[{v'_x}^2+{v'_y}^2+(c\beta-v'_z)^2\big]/2kT}.
    \end{split}
    \end{align}
    \end{widetext}
    The only significant particle distribution is then when $v'_z\sim c\beta$, leading to a highly anisotropic distribution.
    
    In the ISM rest frame the particle distribution is largely isotropic (except for stellar wind and magnetic field effects) while in the spacecraft frame the particle distribution is highly anisotropic and strongly peaked in the spacecraft velocity (relative to the ISM) direction (forward peaked). This has another relevant effect, namely that the ``slant range'' of the particles impacting the spacecraft is greatly increased. We can parameterize the particle distribution in spherical coordinates $(\theta, \phi)$ where $\theta$ is along the z-axis and $\phi$ is in the x-y plane with $\theta=\cos^{-1}(v_z/v)$ and $\phi=\tan^{-1}(v_y/v_x)$. For a planar structure, such as a thin disk, the slant range is
    \begin{align}
        \begin{split}
            R_{\textrm{slant}}= \frac{h}{\sin(\theta)} = \frac{h}{\sqrt{1-\cos^2(\theta)}}= \frac{hv}{\sqrt{v^2-v_z^2}},
        \end{split}
    \end{align}
    where $h$ is the thickness of the material in question. The net effect here is that the disk and any coating on it will appear to be effectively much thicker due to the high flux in the forward direction.
    
     \begin{widetext}
\subsection{ISM Particle Impact Fluxes}
    The particle impact flux $\Gamma_{P-\bm{v}}(\textrm{\# s$^{-1}$ m$^{-2}$})$ on the outward surface ($2\pi$) with outward normal vector $\bm{n}$ is:
    \begin{equation}
        \Gamma_{P-\bm{v}}=\int_{v_x,v_y,v_z}\iint_{2\pi}n_p\Big(v_xf_{\bm{v}}\big(v_x\big),v_yf_{\bm{v}}(v_y),v_zf_{\bm{v}}(v_z)\Big)\bm{n}\dfrac{d\Omega}{2\pi}\mathop{dv_x}\mathop{dv_y}\mathop{dv_z}.
    \end{equation}
    Note that we use $n_p$ rather than $n_p/2$ as we will allow particles going in both directions in the ISM frame in the case of a moving spacecraft which ``overtakes" both particle directions in $z$ from $-\infty$ to $+c\beta$. Consider the front spacecraft case when the spacecraft motion is towards $+z$ as above. In this case, $\bm{n}=(0,0,1)$ and
    \begin{align}
    \begin{split}
        \Gamma_{P-\bm{v}}(z_{\textrm{normal}},z_{\textrm{travel}})&={\int_{-\infty}^{c\beta} \iint_{2\pi}} n_p v_z f_{\bm{v}} (v_z) \dfrac{d\Omega}{2\pi} \mathop{dv_z}=n_p {\int_{-\infty}^{c\beta}}(c\beta - v_z) f(v_z) \mathop{dv_z}\\
        &=n_p c\beta {\int_{-\infty}^{c\beta}}f(v_z)\mathop{dv_z} - n_p {\int_{-\infty}^{c\beta}} v_z f(v_z) \mathop{dv_z}.
    \end{split}
    \end{align}
    Note that for a relativistic mission, $c\beta$ is extremely large compared to the typical particle speeds in the ISM frame. This means that the ``tails'' of the distribution function are extremely small from speed $c\beta$ to infinity. For practical purposes, we can ignore them and integrate to $\infty$:
    \begin{align}
    \begin{split}
        \Gamma_{P-\bm{v}}(z_{\textrm{normal}},z_{\textrm{travel}})\sim n_p c \beta {\int_{-\infty}^{\infty}} f(v_z)\mathop{dv_z} - n_p {\int_{-\infty}^\infty} v_z f(v_z) \mathop{dv_z} = n_p c \beta.
    \end{split}
    \end{align}
    This is correct since
    \begin{align}
    \begin{split}{}
        \int_{-\infty}^{\infty}f(v_z)\mathop{dv_z} = 1\hspace{1mm}\textrm{and}\hspace{1mm}
        \int_{-\infty}^{\infty} v_z f(v_z) \mathop{dv_z} = 0.
    \end{split}
    \end{align}
    We define the RMS particle velocity as
    \begin{align}
    \begin{split}
        \sigma_{vel}=v_{x\textrm{-rms}}=v_{y\textrm{-rms}}=v_{z\textrm{-rms}} = \left[\dfrac{kT}{m}\right]^{1/2}= c\left[\dfrac{kT}{mc^2}\right]^{1/2},
    \end{split}
    \end{align}
    since 
    \begin{equation}
        \int_0^\infty v_x e^{-mv_z^2/2kT} \mathop{dv_x} = \frac{kT}{m} = \sigma_{vel}^2.
    \end{equation}
    For the $x$ and $y$ normal vector surfaces, we integrate from 0 to $\infty$ since the particles hit one side:
    \begin{align}
    \begin{split}
        \Gamma_{P-x}(x_{\textrm{normal}},z_{\textrm{travel}}) ={\int_0^\infty \iint_{2\pi}} n_p v_x f_{\bm{v}} (v_x) \dfrac{d\Omega}{2\pi} \mathop{dv_x}= n_p \left[ \dfrac{m}{2\pi kT} \right]^{1/2} {\int_0^\infty} v_x e^{-mv_z^2/2kT} \mathop{dv_x} = n_p \left[ \dfrac{kT}{2\pi m} \right]^{1/2}.
    \end{split}
    \end{align}
    Therefore, we have
    \begin{equation}
        \Gamma_{P-x}(x_{\textrm{normal}},z_{\textrm{travel}}) = \Gamma_{P-y}(y_\textrm{normal},z_\textrm{travel}) = \dfrac{n_p \sigma_{vel}}{\sqrt{2\pi}}.
    \end{equation}
    We can then define a ``flux vector'' for the case of a high-speed ($c\beta >>\sigma_{vel}$) spacecraft moving along the z-axis:
    \begin{align}
    \begin{split}
        \bm{\Gamma}=(\Gamma_{P-x},\Gamma_{P-y},\Gamma_{P-z})=\left(\dfrac{n_p \sigma_{vel}}{\sqrt{2\pi}} \cos(\phi),\dfrac{n_p \sigma_{vel}}{\sqrt{2\pi}} \sin(\phi),n_p c\beta \right).
    \end{split}
    \end{align}
    \newpage
    
    \begin{figure*}
        \centering
        \begin{tabular}{cc}
              \hspace{-15mm}\includegraphics[width=0.445\textwidth]{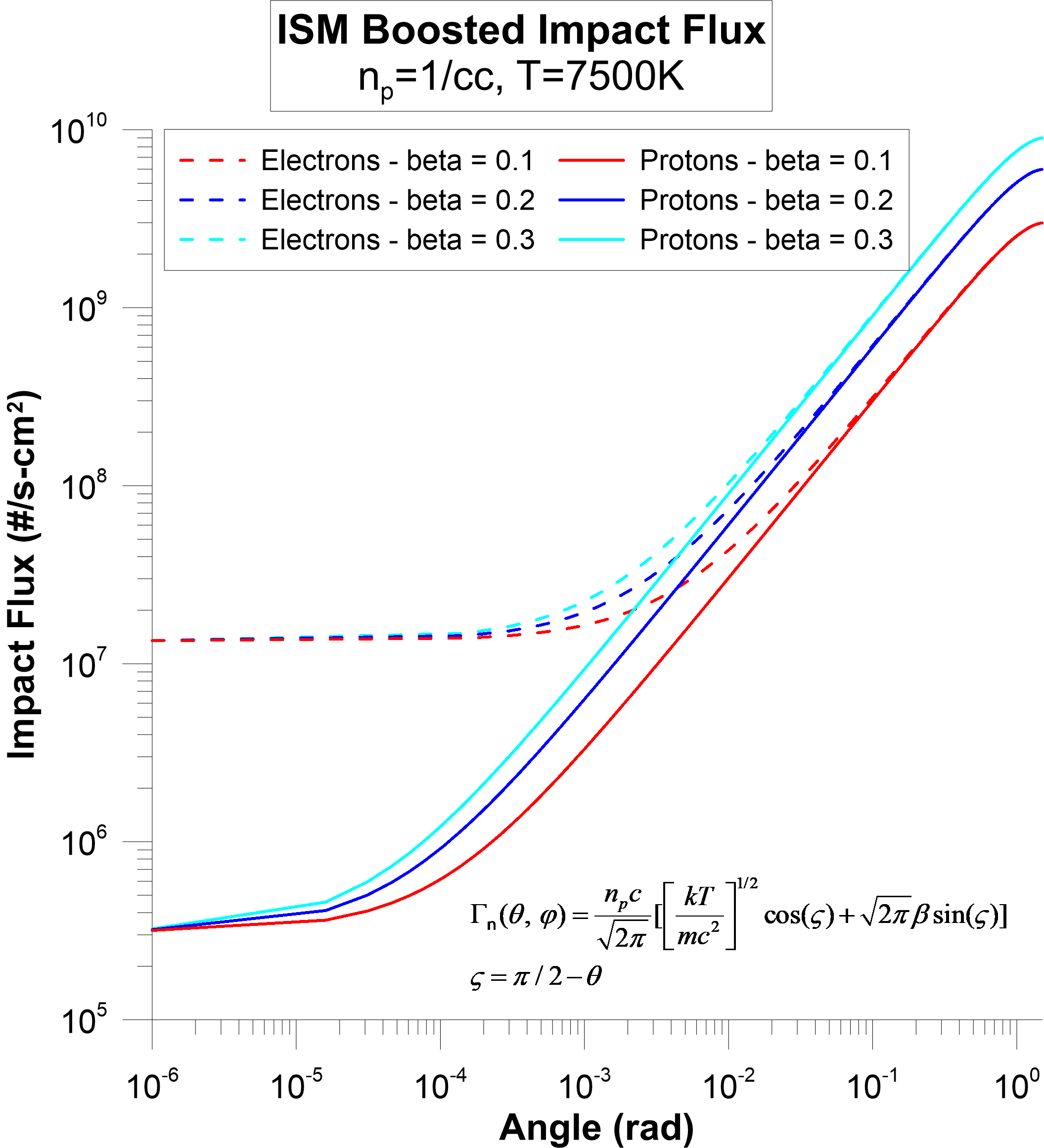} & \includegraphics[width=0.4375\textwidth]{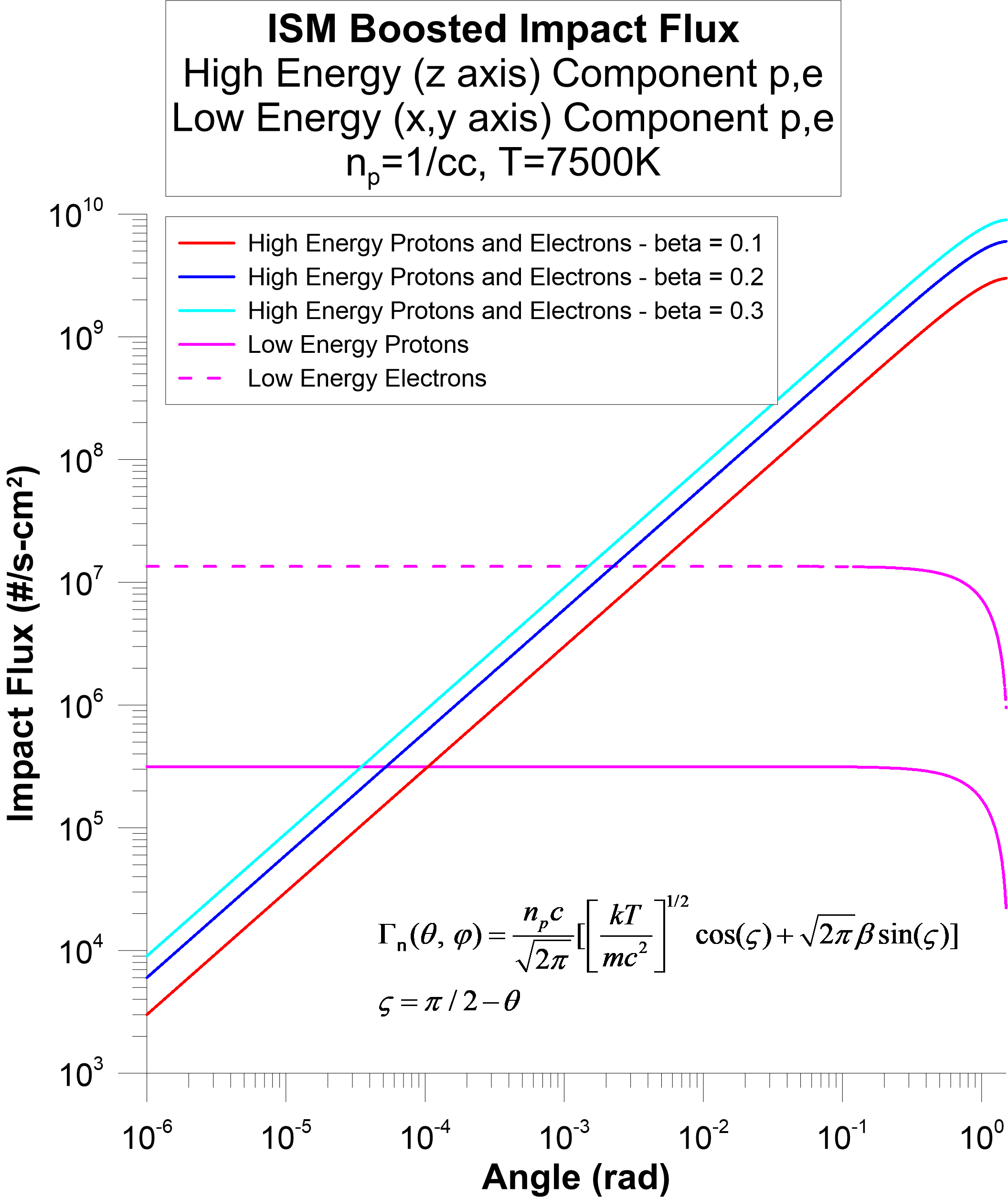}\\
              (a) & (b)
        \end{tabular}
        \caption{(a) Total ISM impact flux as a function of the co-angle $\zeta$ relative to the $x$-$y$ plane (see Equation \ref{eq:coangle}) for protons (solid curves) and electrons (dashed curves) for number density 1 cm$^{-3}$, temperature 7500 K, and spacecraft $\beta$ ranging from 0.1 (red), to 0.2 (blue), to 0.3 (cyan). It can be seen how the boosted impact flux is strongly peaked in the forward direction around $\zeta=0$. (b) ISM impact flux directional components, showing the boosted flux along the $z$-axis, parallel to spacecraft velocity vector, and the lower energy components along the $x$- and $y$-axes perpendicular to the spacecraft velocity vector (magenta curves). These components sum to produce the total ISM impact flux plotted in (a).}
        \label{fig:ISM_impact_flux}
    \end{figure*}
    
    \end{widetext}
    
    For a surface with normal $\bm{n}$ the magnitude of the flux in [\# s$^{-1}$ m$^{-2}$] is $\Gamma_n=\bm{\Gamma}\cdot \bm{n}$. For a surface with normal vector relative to the spacecraft velocity vector (the $z$-axis), we have:
    \begin{equation}
        \bm{n}=[\mathop{\sin(\theta)\cos(\phi)},\mathop{\sin(\theta)\sin(\phi)},\mathop{\cos(\theta)}].
    \end{equation}
    Here, $\theta$ and $\phi$ are the usual spherical coordinate angles relative to the $z$-axis. We thus obtain
    \begin{align}
    \begin{split}
        \hspace{-2mm}\Gamma_n(\theta,\phi)&=n_p\left[\dfrac{\sigma_{vel}}{\sqrt{2\pi}} \sin(\theta)+c\beta \cos(\theta) \right]\\
        &= \dfrac{n_p c}{\sqrt{2\pi}}\left[\left(\dfrac{kT}{mc^2}\right)^{\hspace{-1mm}1/2}\hspace{-3mm}\sin(\theta)+\sqrt{2\pi}\beta\cos(\theta)\right].
    \end{split}
    \end{align}

    Assuming a high speed spacecraft, comparing the surface flux for the surfaces whose normals are along the $z$- vs. $x$-, $y$-axes is useful:
    \begin{align}
    \begin{split}{}
        \dfrac{\Gamma_{P-z}}{\Gamma_{P-x}} &= \dfrac{\Gamma_{P-z}}{\Gamma_{P-y}}=\dfrac{\sqrt{2\pi}c\beta}{\sigma_{vel}}\\&=\sqrt{2\pi} \beta \left(\dfrac{mc^2}{kT}\right)^{1/2}>>1.
    \end{split}
    \end{align}
    We then plot the flux vs. the ``co-angle'' relative to the $x$-$y$ plane, $\zeta=\pi/2-\theta$, in Figure \ref{fig:ISM_impact_flux} as
    \begin{align}
    \begin{split}
        \hspace{-3.075mm}\Gamma_n(\theta,\phi)&=\dfrac{n_pc}{\sqrt{2\pi}}\left[\left(\dfrac{kT}{mc^2}\right)^{\hspace{-1mm}1/2} \hspace{-3mm}\sin(\theta) +\sqrt{2\pi} \beta \cos(\theta)\right]\\
        &=\dfrac{n_pc}{\sqrt{2\pi}}\left[\left(\dfrac{kT}{mc^2}\right)^{\hspace{-1mm}1/2} \hspace{-3mm}\cos(\zeta) +\sqrt{2\pi} \beta \sin(\zeta)\right].
        \label{eq:coangle}
    \end{split}
    \end{align}
    The plots in Figure \ref{fig:ISM_impact_flux} are for incident protons and electrons, as well as in high energy (along $z$-axis and velocity vector, and therefore boosted) and low energy (along $x$- and $y$-axes) components. In order to reduce the particle flux, it is important to minimize the angle $\zeta$ of critical elements exposed or (preferred) using a raised edge at the front (leading edge) surface.
    

    \begin{widetext}{}

\subsection{ISM Particle Angular Distribution}

    Since the spacecraft is moving along the $z$ direction at speeds much greater than the ISM frame particle speed $v$, the speed of the ISM particles in the frame of the spacecraft $v'$ is dominated by the $z$ component, $v'_z$:
    \begin{align}
    \begin{split}
        v'_x=v_x=v'\sin(\theta)\cos(\phi), \hspace{3mm}v'_y=v_y=v'\sin(\theta)\cos(\phi), \hspace{3mm}v'_z=c\beta-v_z=v'\cos(\theta).
    \end{split}
    \end{align}
    We can write a function $g(v',\beta,\theta)$ which is the ISM frame particle speed $v$ as a function of the spacecraft frame particle speed $v'$:
    \begin{align}
    \begin{split}{}
        v=\big[v_x^2+v_y^2+v_z^2\big]^{1/2}=\big[{v'_x}^2+{v'_y}^2+(c\beta-v'_z)^2\big]^{1/2}=\big[{v'}^2\sin^2(\theta)+(c\beta-v'\cos(\theta))^2\big]^{1/2}\equiv g(v',\beta,\theta),
    \end{split}
    \end{align}
    which provides the following convenient shorthand:
    \begin{align}
    \begin{split}{}
        g(v',\beta,\theta)\equiv c\beta\Bigg[1+\bigg(\frac{v'}{c\beta}\bigg)^2-2\bigg(\frac{v'}{c\beta}\bigg)\cos(\theta)\Bigg]^{1/2}.
    \end{split}
    \end{align}
    We can then write the ISM frame particle velocity distribution as
    \begin{align}
    \begin{split}{}
        f(v)&=4\pi v^2 \Big[\frac{m}{2\pi kT}\Big]^{3/2}e^{-mv^2/2kT}\\
        &=4\pi\left[v'^2\sin^2(\theta)+(c\beta-v'\cos(\theta))^2\right]\Big[\frac{m}{2\pi kT}\Big]^{3/2}e^{-m\big[v'^2\sin^2(\theta)+(c\beta-v'\cos(\theta))^2\big]/2kT}.
    \end{split}
    \end{align}
    We can relate the two distribution functions $f(v)$ and $f(v')$ since $f(v)dv=f(v')dv'$:
    \begin{align}
    \begin{split}
        dv'^2&=(dv'_x)^2+(dv'_y)^2+(dv'_z)^2=(dv_x)^2+(dv_y)^2+(dv_z)^2=dv^2\\
        &\rightarrow dv'=dv \rightarrow f(v)=f(v')\equiv f(v'(v,\beta,\theta,\phi)).
    \end{split}
    \end{align}
    We can write the distribution function in the spacecraft frame as above:
    \begin{equation}
        f(v')=f(v)=\frac{4}{\sqrt{\pi}v_{mp}}\Bigg[\frac{v}{v_{mp}}\Bigg]^{1/2}e^{-[v/v_{mp}]^{1/2}}=\frac{4}{\sqrt{\pi}v_{mp}}\Bigg[\frac{g(v',\beta,\theta)}{v_{mp}}\Bigg]^{1/2}e^{-[g(v',\beta,\theta)/v_{mp}]^{1/2}},
    \end{equation}
    where $v_{mp}=\sqrt{2kT/m}$, which satisfies $df(v)/dv=0$. The particle densities $n'_p$ and $n_p$ are the same in both the ISM and spacecraft frames, but the distribution $n(v')$ is not the same in the spacecraft frame:
    \begin{align}
    \begin{split}
        n(v)&\equiv n_p f(v)\\
        n(v')&\equiv n'_p f(v')
    \end{split}
    \end{align}
    \begin{align}
    \begin{split}
        n(v')\mathop{dv'}=n'_p f(v') \mathop{dv'}=n(v) dv=n_p f(v) \mathop{dv}&\rightarrow n'_p=n_p \hspace{2mm}\textrm{as}\hspace{2mm} dv'=dv \hspace{2mm}\textrm{and}\hspace{2mm} f(v')=f(v)\\
        n(v')&=n(v'(v,\beta,\theta,\phi)).
    \end{split}
    \end{align}
    Since $c\beta\gg v \rightarrow v'\sim c\beta$. Therefore, we can write the ISM particle velocity distribution as
    \begin{align}
    \begin{split}
        f(v)&\sim 4\pi c^2\beta^2\big[\sin^2(\theta)+(1-\cos(\theta))^2\big]\Bigg[\frac{m}{2\pi kT}\Bigg]^{3/2}e^{-mc^2\beta^2[\sin^2(\theta)+(1-\cos(\theta))^2]/2kT}\\
        &=4\pi c^2\beta^2\big[2(1-\cos(\theta))\big]\Bigg[\frac{m}{2\pi kT}\Bigg]^{3/2}e^{-mc^2\beta^2[2(1-\cos(\theta))]/2kT}=\frac{16\pi}{c\beta}\sin^2(\theta/2)\Bigg[\frac{E}{\pi kT}\Bigg]^{3/2}e^{-4E\sin^2(\theta/2)/kT},
    \end{split}
    \end{align}
    where $E=mv'^2/2=mc^2\beta^2/2$ is the kinetic energy of impacting particles in the spacecraft frame. 
    \newpage
    \end{widetext}
    Note that the particle impact kinetic energy is dominated by the spacecraft speed and not the particle speed in the ISM reference frame. Therefore, the spacecraft sees nearly mono-energetic impacts for a given species, as the ISM speeds are much smaller than the spacecraft speed. However, the angular distribution function is highly peaked in the forward direction, near $\theta=0$. This allows the possibility of a slightly raised edge shield to block the boosted ISM (like a rain guard on a car). For example, in the above analysis, $E/kT = 2.9\times10^7$ for 19 MeV protons and $1.5\times10^4$ for 10 keV electrons, assuming a spacecraft $\beta=0.2$ and $T=7500$ K for the ISM with protons, and the angular distribution peaks at about $10^{-4}$ radians. 
    
\subsection{Raised Edge Shield}
    
    For the example above the proton angular distribution for $\beta=0.2$ peaks at $\theta\sim10^{-4}$ rad with a steep decrease at $10^{-3}$ rad. A forward raised shield for a 10 cm diameter disk that was raised $10^{-3}$ rad would be $10^{-3}\times 100$ mm = 0.1 mm high. A $\beta=0.2$ proton is 19 MeV and has a penetrating range of about 2.2 mm in Si (see next section). Assuming a shield all around the perimeter of a 10 cm diameter (round) wafer, this would have a mass of about 0.2 g assuming the shield is made of Si and is 0.1 mm high and 3mm thick. This mass would be potentially acceptable for a 10 cm diameter wafer design. If the orientation of the spacecraft could be assured during flight, as discussed extensively in Section 10 of \citep{lubin_roadmap_2016}, then the mass could be reduced to below 0.1 g as a ``straight edge forward shield'' could be used. Spacecraft attitude oscillations caused by interactions with ISM magnetic fields are discussed extensively in \citep{hoang_interaction_2017}, wherein the authors find that impacts with ISM particles will positively charge the spacecraft to equilibrium (defined when the surface potential of the charged spacecraft no longer allows electrons to be ejected) within 0.1\% of the path length to $\alpha$ Centauri. Since the spacecraft charging only occurs locally near the leading edge and is limited by the implantation depth of incident particles, the spacecraft will develop an electric dipole moment which interacts with the ISM magnetic field so as to produce oscillations about the axis perpendicular to the velocity vector with a period of $\sim0.5$ hours \citep{hoang_electromagnetic_2017}. This would have the effect of periodically exposing larger surface areas of the spacecraft to the ISM flux, thus rendering a raised edge shield ineffective. However, such a spacecraft would necessarily have on-board attitude control systems, as discussed in \citep{lubin_roadmap_2016}, such as photon thrusters or small field emission type electric ion thrusters, the capabilities of which would be driven by the torque required to modify initial trajectories and to mitigate dust grain impacts and magnetic/electric field trajectory perturbations. Photon thrusters for example, like laser diodes or LED's, can achieve $10^{-10}$ Nm-s of torque impulse, which is enough to mitigate dust grain impact attitude perturbations. Small ion thrusters could yield $\>5000\times$ more thrust per unit power if needed \citep{lubin_roadmap_2016}. As discussed in \citep{hoang_interaction_2017}, spacecraft attitude perturbations may be further mitigated in a number of ways, including applying a fast rotation to the spacecraft about it's central axis (through the disk's center) so as to increase its angular momentum, launching the spacecraft with a net negative charge, or optimizing the spacecraft geometry to minimize its forward-facing surface area (like a needle) and applying a fast rotation about its long axis \citep{hoang_interaction_2017}.
    
    


\section{Penetration Depths}
\label{section:pendepths}

\subsection{Electron Penetration}

    The $dE/dx$ and range of electrons in materials can be computed using Bethe-Bloch theory of electron ionization, though deviations for real materials have been measured and are included here. The range of 1-10 KeV the range of electrons is well characterized by the following equation \citep{feldman_range_1960}:
    \begin{equation}
        R=25\frac{A}{\rho Z^{n/2}}E^n,
    \end{equation}
    where $n=1.2/(1-0.29\log(z))$, $R$ is the range in nanometers, $E$ is in keV, $\rho$ is the density (g cm$^{-3}$), $Z$ is the atomic number for elements or electrons per molecue, and $A$ is the atomic mass (g mole$^{-1}$).
    
    For Si we have $Z=14$, $A=28.09$ g mole$^{-1}$, and $\rho=2.33$ g cm$^{-3}$. The energy loss can be expressed from the Bethe equations as:
    \begin{equation}
        -\frac{dE}{dx}=7.8\times10^{-3}\frac{\rho Z}{AE}\ln\bigg(\frac{1.16E}{I}\bigg),
    \end{equation}
    where $I$ is the excitation energy averaged over energy levels in the atom/molecule. $dE/dx$ has units of keV nm$^{-1}$ here. An approximation is $I=13.5\cdot Z$(eV), but a better approximation is \citep{kramers_xciii_1923}:
    \begin{align}
        \begin{split}
            I&\sim 19.0\hspace{1mm}\textrm{eV}; \hspace{1mm}Z=1\\
            I&\sim(11.2+11.7\times Z) \hspace{1mm}\textrm{eV}; \hspace{1mm} Z=2 \hspace{1mm} \textrm{through} \hspace{1mm} 13\\
            I&\sim(52.8+8.71\times Z) \hspace{1mm} \textrm{eV}; \hspace{1mm} Z>13.
            \label{I}
        \end{split}
    \end{align}
    Therefore, for Si ($z=14$), $I\sim 175$ eV $=0.175$ keV.

\subsection{Proton Penetration}

    Note that for thin spacecraft, such as the waferscale spacecraft we propose in our NASA Starlight (DEEP-IN and DEIS) programs and the Breakthrough Starshot program, the protons can penetrate through both sides and hence the bombardment rate is effectively $2\Gamma_p\sim 1.3\times10^5$ hits cm$^{-2}$ s$^{-1}$. As we discussed in our previous ``roadmap'' paper \citep{lubin_roadmap_2016} the leading edge of the spacecraft that is hit by ISM protons can be protected with a thin boundary edge ``guard ring'' that is about 1 cm wide (but very thin). However the effect of ``transverse'' bombardment of the spacecraft is much more difficult to deal with. Comparing the proton ISM transverse bombardment to the galactic proton cosmic ray bombardment we see the ISM effects are vastly larger, of order $10^6$ times larger, though the energy of the protons is nearly mono-energetic at 19 MeV for $\beta=0.2$. The problem is the shielding required to stop a 19 MeV proton is roughly 4 mm of water equivalent or 2.2 mm in Si. Ideally we would develop electronics that could withstand the ISM transverse bombardment without the need of shielding. A simple calculation shows that over a 30 year mission ($\sim10^9$ s) the number of hits would be about $10^{14}$ protons cm$^{-2}$  (at 19 MeV), or per 1$\times$1 micron cell (typical of a semiconductor unit cell) the number of hits would be $10^6$ $\mu$m$^{-2}$. 
    
    We can use the relativistic Bethe-Bloch energy loss equation for calculating charged particle energy deposition due to collisional ionization. Note that the energy loss $dE/dx$ for non relativistic particles is roughly proportional to $z^2/\beta^2$ and hence slower and higher $Z$ particles (He and beyond) deposit much greater energy per unit length than protons for the same $\beta$. This also means that as the particles slow down during penetration into the material they increase their energy loss per unit length $dE/dx$ as they slow down so that maximum $dE/dx$ is near the end (there is a finite cutoff) of the range. For galactic cosmic rays, $dE/dx$ is such a small portion of the initial energy that $dE/dx$ is nearly constant for the spacecraft materials and thicknesses we are considering as the galactic cosmic rays largely exit the spacecraft with nearly the same exiting $\beta$ as the entering $\beta$. For the ``boosted'' ISM interactions with protons the particles exit with significant affect (i.e. large energy loss in the spacecraft frame - in some cases they stop inside the spacecraft for lower $\beta$ ($<0.1$) spacecraft) while for larger $\beta$ ($>0.2$) they pass through with only modest energy loss in the spacecraft frame (i.e. nearly unaffected) while ISM electrons are stopped in the initial thin layer of the spacecraft for virtually all cases. However, as discussed above, if we use a raised edge shield we can stop almost all of the boosted ISM protons and stop virtually all of the electrons.
    
    The Bethe-Bloch relativistic energy loss formula is:
    \begin{equation}
        -\frac{dE}{dx}=\frac{4\pi k^2 Z^2 e^4 n}{mc^2\beta^2}\Bigg[\ln\Bigg(\frac{2mc^2\beta^2}{I(1-\beta^2)}\Bigg)-\beta^2\Bigg],
    \end{equation}
    where $k$ is the Boltzmann constant, $Z$ is the atomic number of the incoming particle, $n$ is the electrons per unit volume in the material, $m$ is the electron mass, and $I$ is the mean excitation energy in the material. We approximate $I$ as in Equation \ref{I}.
    
    For composite materials the weighted densities must be used to compute the effective mean excitation energy for the material $I_{\textrm{total}}$ as follows:
    \begin{equation}
        \ln(I_{\textrm{total}})=\frac{\sum_i N_i Z_i \ln I_i}{\sum_i N_i Z_i}=\sum_i \frac{N_i Z_i}{n}\ln I_i,    
    \end{equation}
    where $n=\sum_i N_i Z_i$ is the total number of electrons and $N_i$ is the number of electrons of species $Z_i$.
    
\begin{figure*}
        \centering
        \begin{tabular}{cc}
             \includegraphics[width=0.45\textwidth]{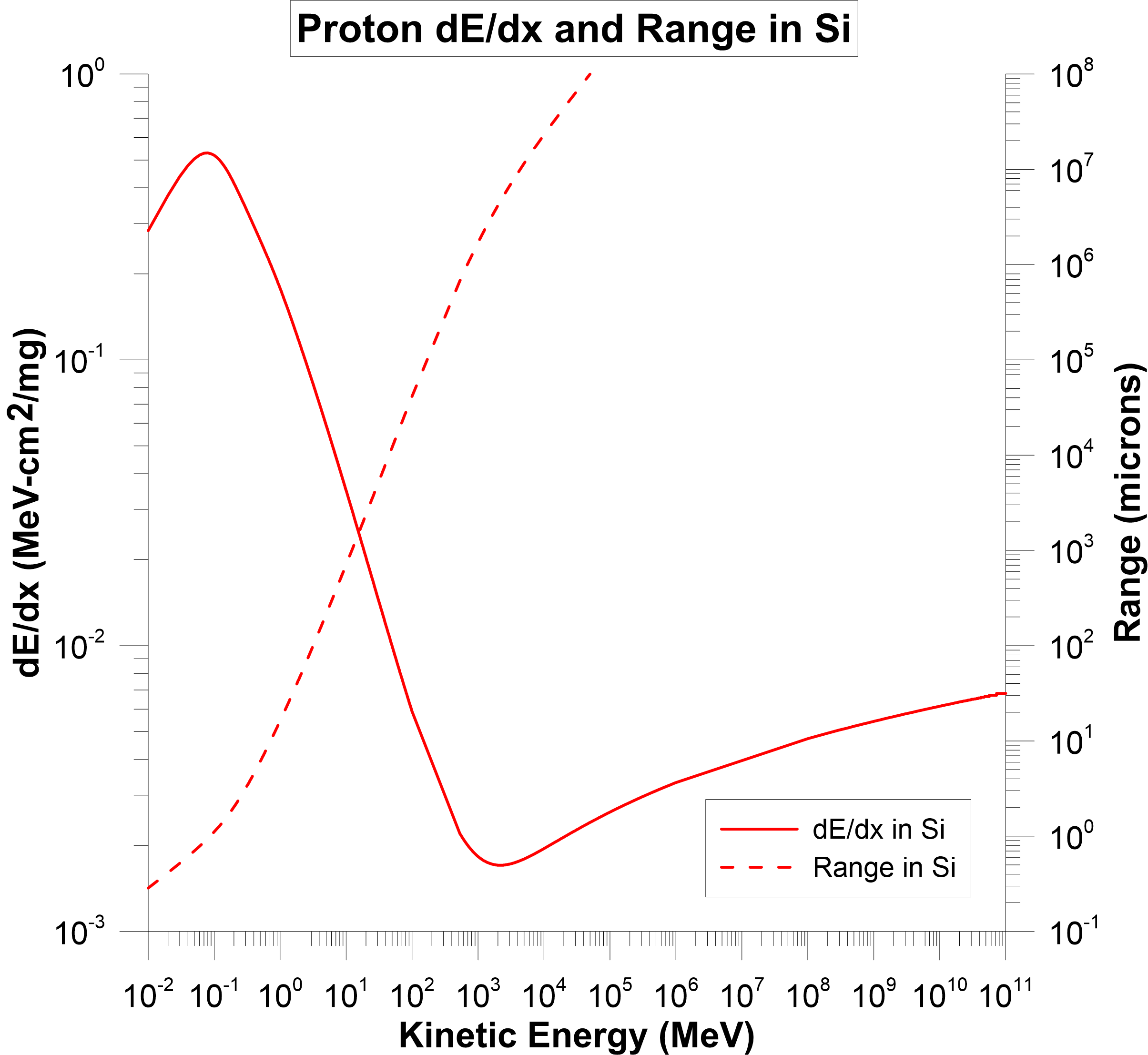} & \includegraphics[width=0.45\textwidth]{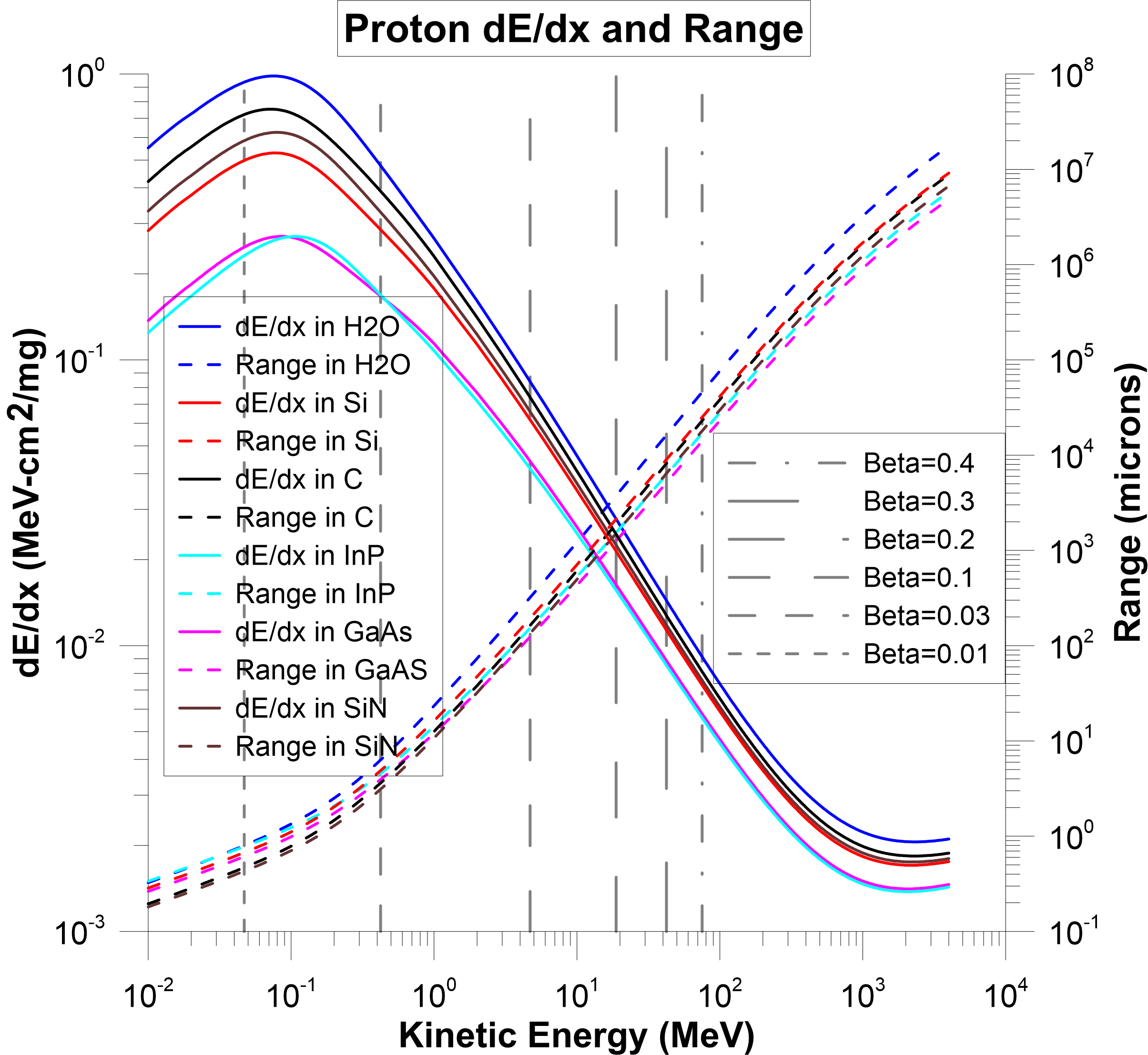} \\
             (a) & (b) \\
             \\
             \includegraphics[width=0.45\textwidth]{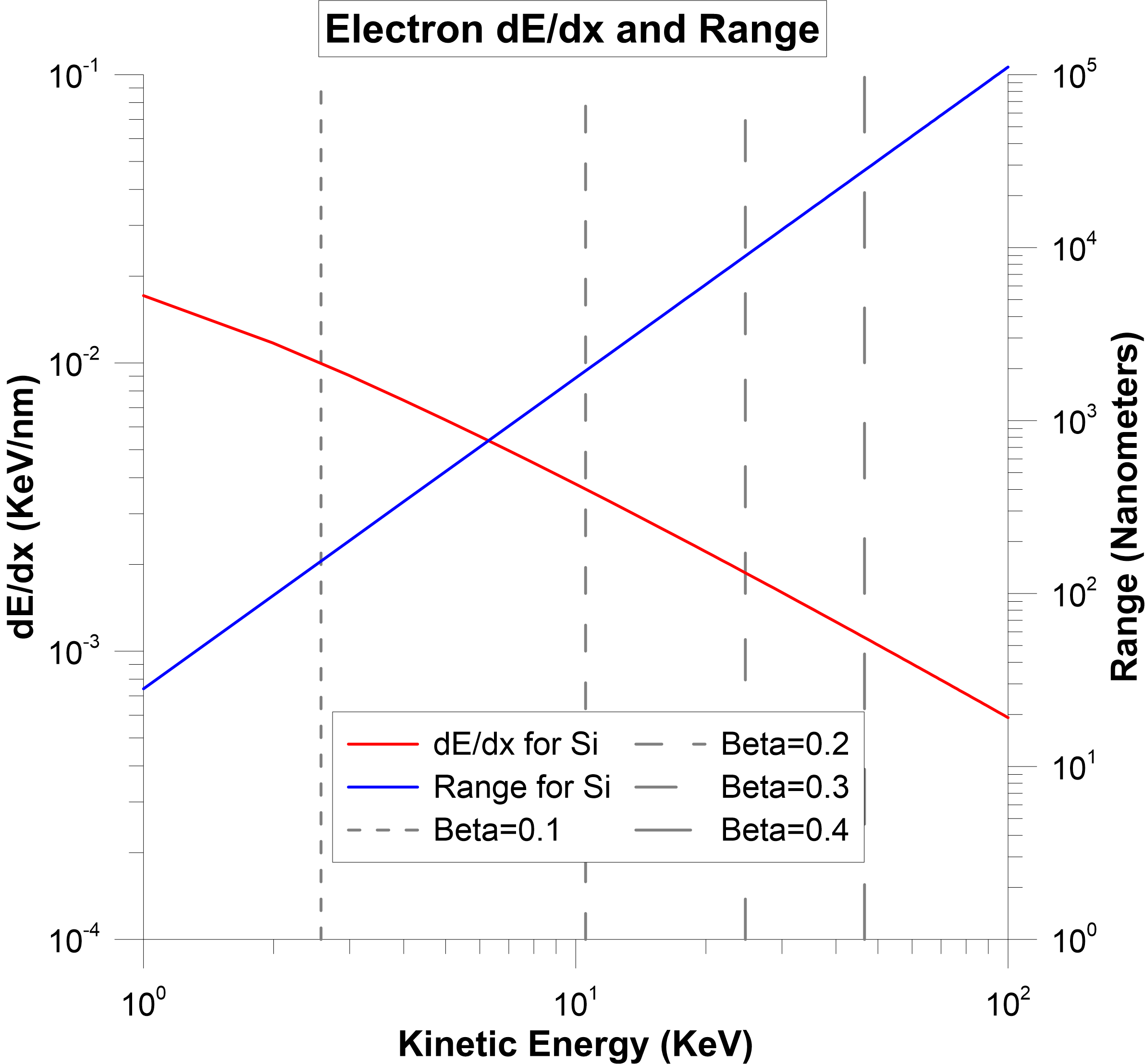} & \includegraphics[width=0.45\textwidth]{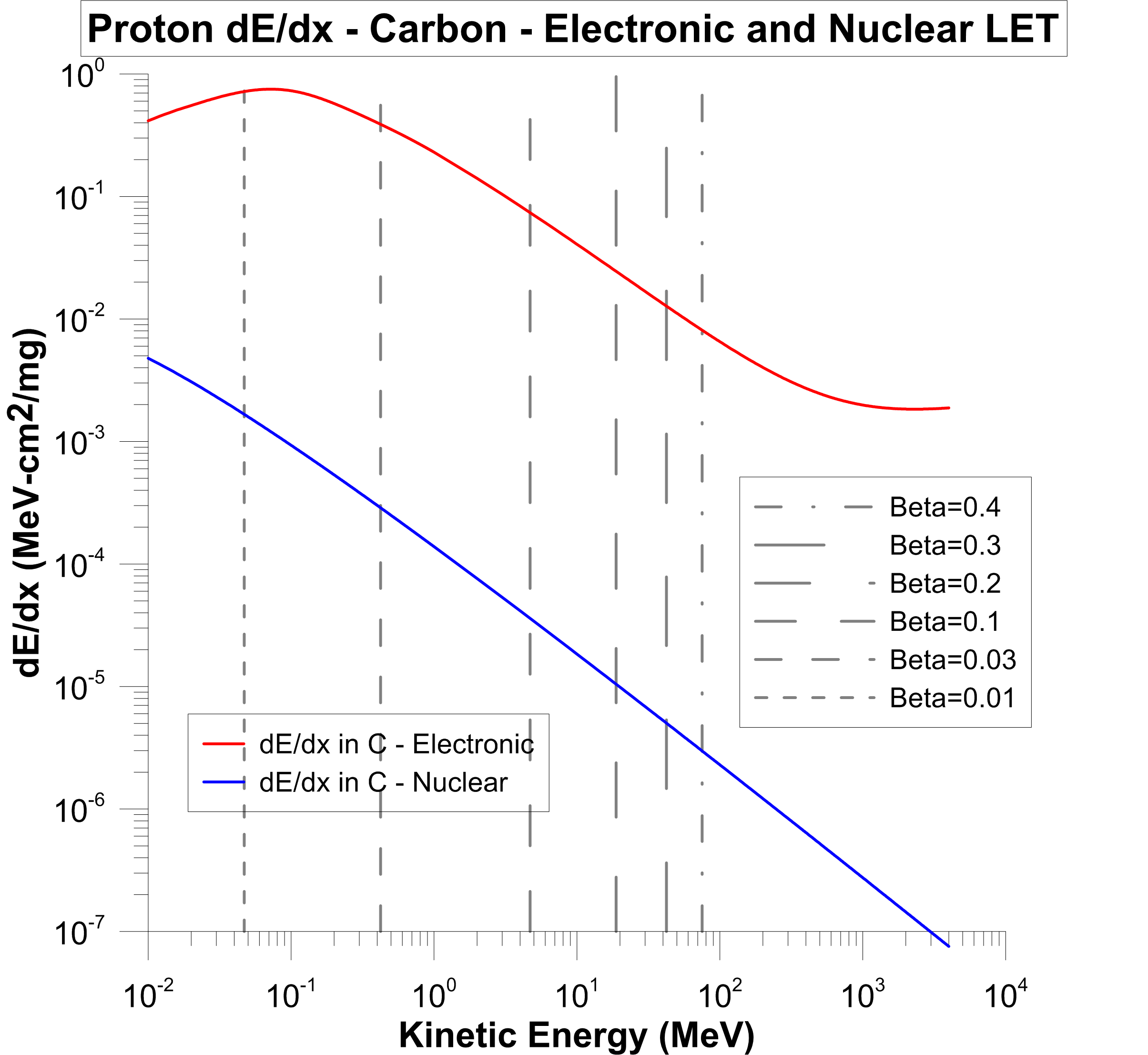} \\
             (c) & (d)
        \end{tabular}
        \caption{(a) Energy loss per unit penetration distance (also referred to as the stopping power) as well as penetration range given as a function of kinetic energy for protons in silicon. Calculated using the NIST PSTAR database, which includes both nuclear and electronic stopping components \citep{berger_stopping-power_2009}. See (d) for a comparison of the electronic and nuclear components. (b) Similar plot showing the stopping power and range of protons incident upon a variety of target materials. The vertical grey lines indicate the proton energies for various specific spacecraft $\beta$, as shown in the legend. (c) Plot of the stopping power and range for electrons incident upon a silicon target. Similarly, the vertical grey lines indicate the electron kinetic energy for specific spacecraft $\beta$. (d) Electronic and nuclear components of the stopping power for protons incident upon a carbon target. It can be seen how the electronic component of the stopping power dominates over the nuclear component for all energies of interest.}
        \label{fig:dEdx_range_vs_energy}
    \end{figure*}

    Using this we compute proton energy loss $dE/dx$ and range in Si, SiN, GaAs, InP, and H$_2$O versus energy, plotted in Figure \ref{fig:dEdx_range_vs_energy}, including both electronic and nuclear linear energy transfer (LET) mechanisms courtesy of the NIST PSTAR database \citep{berger_stopping-power_2009} and the ``Energy vs. LET. vs. Range calculator'' devloped by Vladimir Zajic of Brookhaven National Laboratory \citep{zajic_energy_2002}. The electronic energy loss dominates over the nuclear interaction energy loss in all cases.
    
    H$_2$O was chosen as an example of a low $Z$ material to compare to Si. For 19 MeV protons in Si $dE/dx$ = 21 MeV g$^{-1}$ cm$^{-2}$ or 0.021 MeV mg$^{-1}$ cm$^{-2}$ and a range of 2.2 mm. With a Si density of 2.33 g cm$^{-3}$ we have an area density per micron thickness of 0.233 mg cm$^{-2}\mu$m$^{-1}$ = $2.3\times10^{-7}$ kg cm$^{-2}\mu$m$^{-1}$. Using $dE/dx$ of a 19 MeV proton of 0.021 MeV mg$^{-1}$ cm$^{-2}$ we would have $dE/dx$ equivalent of 4.9 keV$\mu$m$^{-1}$. It is possible to build thinned Si circuit elements with a thickness of less than 1 micron and hence this allows us an interesting comparison of energy loss for thinned Si circuits.  For such thin circuits we use the two sided ISM transverse proton bombardment rate of $2\Gamma_p\sim1.3\times10^5$ hits  cm$^{-2} $ s$^{-1}$. This gives an ISM proton deposition rate per micron thickness of Si of  $1.3\times10^5$ hits cm$^{-2}$ s$^{-1}\times4.9$ keV$\mu$m$^{-1}$ = 0.1 nW cm$^{-2}\mu$m$^{-1}$. Normally device irradiation is given in ``rads'' (radiation absorbed dose) where 1 rad = 0.01 J kg$^{-1}$ or 100 erg g$^{-1}$ or in Grays (1 Gy = 100 rad = 1 J kg$^{-1}$).  The Si energy loss from the ``transverse boosted'' ISM protons of 0.1 nW cm$^{-2} \mu$m$^{-1}$ gives a ``rad'' rate of 0.1 nW cm$^{-2} \mu$m$^{-1}\cdot(2.3\times10^{-7}$kg)$^{-1}$ cm$^{-2} \mu$m$^{-1}=4.3\times10^{-4}$ W kg$^{-1} \rightarrow 4.3\times10^{-4}$ W kg$^{-1}$(0.01 W s)$^{-1}$ kg$^{-1}$= 0.04 rad s$^{-1}$ = 1.2 Mrad yr$^{-1}$. Over a 30 year mission this would be a total integrated dose of about 35 Mrad. Note that these doses are relatively insensitive to the thickness of the Si used, since $dE/dx$ is approximately linear over the device thicknesses we envision,  though we are tending toward thinner devices to allow ``3D'' circuit redundancy and higher effective devices density.

\subsection{Nuclear and Electronic Impact Ionization Losses}
    
    There are two terms in the impact ionization for radiation losses. These are the nuclear and electronic ionization terms. Each of these contributes to the ionization. The electronic loss term generally completely dominates the loss. This is important for long term material damage. We compute each term for carbon as an example. Other materials behave in a similar fashion. For thin materials this is particularly important as electrons will be ejected in each collision. This will cause the material to charge positive. This will then cause electrons from the ISM to be attracted (and protons to be repelled) and hence there will be a space charge limit. 

\section{Production of Photons by Charged Particle Impacts}
\label{section:brem}

    The energy of particle impacts goes almost completely into heat (phonon modes) with a small fraction ($<1$\%) of energy going into photons.  The production of photons by the bombardment of the ISM charged particles (e, p, He, etc.) on the spacecraft is caused by two basic mechanisms. One is photon production via bremsstrahlung (continuum photons) and the production of photons via inner shell excitation and de-excitation (characteristic photons). Virtually all of the bremsstrahlung comes from the ISM charged particles interacting with the electrons in the material rather than the nuclei due to the much heavier mass, and hence much smaller acceleration, of the nuclei. The theoretical analysis goes back nearly 100 years with an intense amount of work going into the analysis and subsequent design of x-ray tubes for medical and industrial applications. Work by Kramers, Bethe, and Heitler and many others in the 1920's and 1930's has led to a combination of analytic and empirical analyses \citep{kramers_xciii_1923}. A large part of the complexity of photon production is due to the extremely small photon ``channel'' compared to the phonon ``channel'' with the details of the material being important. For non-relativistic impacts, the photon angular distribution is approximately isotropic while for relativistic impacts the photon angular distribution becomes peaked in the forward direction. With each impact the bombarding particle loses energy and slows down and both the spectral shape and angular distribution function change. A proper analysis uses a Monte Carlo analysis to model successive impacts. The two primary impacting species are electrons and protons with a much smaller amount of helium. The electron impacts are very similar to that in x-ray tubes with a large amount of analysis and data to draw from. For proton impacts the primary photon production channel is via impacts with electrons in the material which then impact other electrons and essentially form multiple cascades. Since the proton is $\sim2000$ times more massive than the electrons, the maximum speed of the impacted electrons is twice that of the incident non-relativistic protons.
    
\subsection{Bremsstrahlung Production via Electron Impacts}

    The kinetic energy of an impacting electron (for $\beta<<1$) is given by:
    \begin{equation}
        E_0=m_e(\gamma-1)c^2\sim \frac{1}{2}m_e c^2\beta^2,
    \end{equation}{}where the energy of the bremsstrahlung photon produced, $E_\gamma$, is less than or equal to $E_0$.
    
    A simplified analysis by Kramers \citep{kramers_xciii_1923} gives an approximation for the number of photons per electron impact:
    
    \begin{align}
    \begin{split}
        \frac{dN_\gamma(\#\gamma)}{dE_\gamma}&=a(Z,E_0,E_\gamma)\bigg(\frac{E_0}{E_\gamma}-1\bigg)\\
        &=\frac{a(Z,E_0,E_\gamma)}{E_\gamma}(E_0-E_\gamma),
    \end{split}{}
    \end{align}{}where $a(Z,E_0,E_\gamma)$ is a proportionality constant depending on the material $Z$, electron energy $E_0$, and photon energy $E_\gamma$. To first order, $a\sim Z$ and is independent of $E_0$ and $E_\gamma$, $a(Z,E_0,E_\gamma)\rightarrow a(Z)$. Under this assumption we can compute the total number of bremsstrahlung photons and the total energy. The total number of photons from $E_\gamma=0$ to $E_\gamma=E_0$ is given by:
    \begin{align}
    \begin{split}
        N_\gamma&=\int_0^{E_0} \frac{dN_\gamma}{dE_\gamma}\mathop{dE_\gamma}\\
        &=a(Z)[E_0\log(E_\gamma)-E_\gamma]\Big|_{E_\gamma=0}^{E_\gamma=E_0}\rightarrow\infty,
    \end{split}{}
    \end{align}{}which diverges to infinity at $E_\gamma=0$. The total energy of photons from $E_\gamma=0$ to $E_0$ is finite and given by:
    \begin{align}
    \begin{split}
        E_\textrm{tot}&=\int_0^{E_0}E_\gamma \frac{dN_\gamma}{dE_\gamma}\mathop{dE_\gamma}\\
        &=a(Z)\int_0^{E_0}(E_0-E_\gamma)\mathop{dE_\gamma}\\
        &=a(Z)\bigg[E_0 E_\gamma-\frac{1}{2}E_\gamma^2\bigg]\bigg|_{E_\gamma=0}^{E_\gamma=E_0}=\frac{1}{2}a(Z)E_0^2.
    \end{split}{}
    \end{align}{}This can be interpreted as the area under the triangle of base $a(Z)E_0$ and height $E_0$. The number of photons diverges, but the total energy of photons is finite, as it must be.
    
    An empirical fit to photon bremsstrahlung per unit solid angle and electron current via electron impacts from x-ray tubes for various materials is given approximately by \citep{kramers_xciii_1923}. In this approximation,
    \begin{equation}
        \frac{dN_\gamma}{\mathop{dE_\gamma}\mathop{d\Omega}\mathop{dt}}=iKZ^n\bigg(\frac{E_0}{E_\gamma}-1\bigg)^x,
    \end{equation}{}where $i$ is the electron current, $t$ is the irradiation time, and $K=1.35\times10^9$ $\gamma$ s$^{-1}$ sr$^{-1}$ mA$^{-1}$ keV$^{-1}$ is an empirically determined constant, $n=1$, and $x=1.109-0.00435Z+0.00175E_0$. This is assuming an isotropic distribution (reasonably valid for non-relativistic electrons).
    
    The total number of photons from $E_\gamma=0$ to $E_0$ is thus given by:
    \begin{align}
    \begin{split}
        N_\gamma&=\int_0^{E_0}\frac{dN_\gamma}{\mathop{dE_\gamma}\mathop{d\Omega}\mathop{dt}}\mathop{dE_\gamma}\mathop{d\Omega}\mathop{dt}\\
        &=\int_0^{E_0}iKZ^n\bigg(\frac{E_0}{E_\gamma}-1\bigg)^x\mathop{dE_\gamma}\mathop{d\Omega}\mathop{dt}\\
        &=4\pi itKZ^n\int_0^{E_0}\bigg(\frac{E_0}{E_\gamma}-1\bigg)^x\mathop{dE_\gamma}\\
        &=4\pi itKZ^n\big[E_0\log(E_\gamma)-E_\gamma\big]\bigg|_{E_\gamma=0}^{E_\gamma=E_0}\rightarrow\infty,
    \end{split}{}
    \end{align}{}which diverges to infinity at $E_\gamma=0$. This is valid if we assume $x=1$, which is a reasonable approximation. The general solution for $x\neq1$ involves a hypergeometric function or an incomplete beta function. The total energy of photons from $E_\gamma=0$ to $E_0$ is thus finite and given by:
    \begin{align}
    \begin{split}
        E_\textrm{tot}&=\int_0^{E_0}E_\gamma\frac{dN_\gamma}{\mathop{dE_\gamma}\mathop{d\Omega}\mathop{dt}}\mathop{dE_\gamma}\mathop{d\Omega}\mathop{dt}\\
        &=\int_0^{E_0}iKZ^n E_\gamma\bigg(\frac{E_0}{E_\gamma}-1\bigg)^x\mathop{dE_\gamma}\mathop{d\Omega}\mathop{dt}\\
        &=4\pi itKZ^n\int_0^{E_0}E_\gamma\bigg(\frac{E_0}{E_\gamma}-1\bigg)^x\mathop{dE_\gamma}\\
        &=4\pi itKZ^n\bigg[E_0 E_\gamma-\frac{1}{2}E_\gamma^2\bigg]\bigg|_{E_\gamma=0}^{E_\gamma=E_0}\\
        &=2\pi itKZ^nE_0^2.
    \end{split}{}
    \end{align}{}Note that $E_0$ above and the total energy of photons are both in keV and the current $i$ is in mA. This can be interpreted as the area under the triangle of base $4\pi itKZ^nE_0$ and height $E_0$.
    
    The efficiency $\epsilon$ of conversion of electron kinetic energy to photon energy is:
    \begin{align}
    \begin{split}
        \epsilon&=\frac{\textrm{total photon energy (in J)}}{\textrm{total electron kinetic energy (in J)}}\\
        &=\frac{4\pi itKZ^n(E_0^2/2)(1.6\times10^{-16}\textrm{J keV$^{-1}$})}{itV_0/1000}\\
        &=10^3(1.6\times10^{-16})\frac{4\pi KZ^n E_0^2}{2000E_0}\\
        &\sim1.36\times10^{-6}Z^nE_0,
    \end{split}{}
    \end{align}{}where $V_0$ (volts) is the equivalent acceleration voltage of the electrons, $V_0=1000\times E_0$ (keV). In the above calculation the term $\frac{1}{2}E_0$ is the energy bandwidth in keV and the photon energy in J is $E_01.6\times10^{-16}$ J keV$^{-1}$ and the photon energy in J is $E_01.6\times10^{-16}$ J keV$^{-1}$. The conclusion is that only a very small fraction ($\sim10^{-4}$) of incident electron kinetic energy goes into photon production. The vast majority goes into heat (phonons).
    
    We can also specify the number of photons per incident electron as:
    \begin{equation}
        \frac{dN_{\gamma/e}}{dE_\gamma\mathop{d\Omega}}=\kappa Z^n\bigg(\frac{E_0}{E_\gamma}-1\bigg)^x,
    \end{equation}{}where $\kappa=1.35\times10^9/6.25\times10^{15}$ $\gamma$ e$^{-1}$ sr$^{-1}$ keV$^{-1}$=2.15$\times19^{-7}$ $\gamma$ e$^{-1}$ sr$^{-1}$ keV$^{-1}$, $n=1$, and as before, $x=1.109-0.00435Z+0.00175E_0$. Note that 1 mA-s $\sim10^{-3}$ C $=(10^{-3}\textrm{C})/(1.60\times10^{-19}\hspace{2mm}\textrm{C/e$^-$})=6.25\times10^{15}\hspace{2mm}\textrm{e$^-$}$.
    
    We can compute the total photon bremsstrahlung energy as before. As discussed above, we can analytically solve for the total energy if we approximate $x=1$:
    \begin{align}
    \begin{split}
        E_\textrm{tot}&=\int_0^{E_0}E_\gamma\frac{dN_{\gamma/e}}{\mathop{dE_\gamma}\mathop{d\Omega}}\mathop{dE_\gamma}\mathop{d\Omega}\\
        &=\int_0^{E_0}\kappa Z^n\bigg(\frac{E_0}{E_\gamma}-1\bigg)^x\mathop{dE_\gamma}\mathop{d\Omega}\\
        &=4\pi\kappa Z^n\int_0^{E_0}E_\gamma\bigg(\frac{E_0}{E_\gamma}-1\bigg)^x\mathop{dE_\gamma}\\
        &=4\pi\kappa Z^n\bigg[E_0E_\gamma-\frac{1}{2}E_\gamma^2\bigg]\bigg|_{E_\gamma=0}^{E_\gamma=E_0}\\
        &=2\pi\kappa Z^n E_0^2.
    \end{split}{}
    \end{align}{}
    
    A similar calculation as in (49) shows that the efficiency of conversion of electron energy to total photon energy is the same as before, as it must be.
    
    \subsection{X-Ray Production and Penetration}

    At 0.2c, the protons are at about 19 MeV, and the electrons are at about 10 keV. At 0.3c, these values are 43 MeV and 22 keV respectively. Comparing to a common dental x-ray, we see the electron bombardment of the forward edge can produce comparable x-ray energies. Assuming an ISM proton and electron density of $n=0.2$ protons cm$^{-3}$ we get $\Gamma_p$ (impacts s$^{-1}$ cm$^{-2}$) $= nv =0.2\beta c=6\times10^9 \beta$ (p,e s$^{-1}$ cm$^{-2}$) $\sim 0.10 \beta$ nA cm$^{-2}$. This is very small compared to normal x-ray tubes where the currents (for medical use) are typically several hundred mA for 0.1-1 s at 50-100 keV. To place this in perspective, this medical x-ray is roughly equivalent (in integrated current) to our front edge current density of 0.10$\beta$ nA cm$^{-2}$ for 3 years at $\beta=0.2$ for a 1 cm$^2$ cross section, but the actual x-ray exposure in our case is much lower since our electron energy is much lower (about 10 keV) and the x-ray production in our case is largely bremsstrahlung rather than an inner K shell line.
    
    For both electrons and protons the x-ray production is through interaction with the material electrons. There are two basic mechanisms for photon (x-ray) production, namely through direct bremsstrahlung and through inner shell electron interactions. The latter is how x-ray tubes are optimized. Hence, the use of high $Z$ targets allows high energy inner shell transitions to produce high energy x-rays. There are a large number of inner shell transitions in materials and certainly many in the keV to tens of keV range. Typically more than 99\% of the electron bombardment energy is dissipated as heat rather than x-rays. In the case of protons, being about $2000\times$ more massive than the electrons, the maximum electron recoil speed is roughly twice that of the impacting proton. 
    
    Just as in the case of charged particle interaction there are loss mechanisms as the x-rays interact with the material. Since the electrons and protons are not highly relativistic, the bremsstrahlung is largely isotropic with some forward scattering enhancement. The inner shell x-ray production is very isotropic. This is good for us as the x-rays produced will essentially all be in the forward edge and will mostly leave isotropically with little interaction with the spacecraft from geometry. The penetration depth of x-rays is a very strong function of energy with higher energies being much more penetrating. X-ray interactions are largely through a combination of Compton scattering, Rayleigh scattering, electron elastic and inelastic collision, Auger and excitation of inner shell electrons.
    
    Unlike protons and electrons which have relatively short range and large $dE/dx$, x-rays have much larger range and consequently lower $dE/dx$. While the electrons and protons do not penetrate the outer edge of the spacecraft the x-rays can easily do so. Luckily the production rate of x-rays is relatively small, is largely isotropic  and the energy deposition $dE/dx$ is small. In general, Si devices can be engineered to be quite insensitive to x-rays so this does not look like it will be an issue. X-ray attenuation is well approximated by an exponential decrease in intensity ($\gamma$ s$^{-1}$ cm$^{-2}$) vs penetration into materials. This is not exact as there is also energy dissipation as well as scattering off the material electrons.
    
\begin{figure*}
        \centering
        \begin{tabular}{cc}
             \hspace{-10mm}\includegraphics[width=0.45\textwidth]{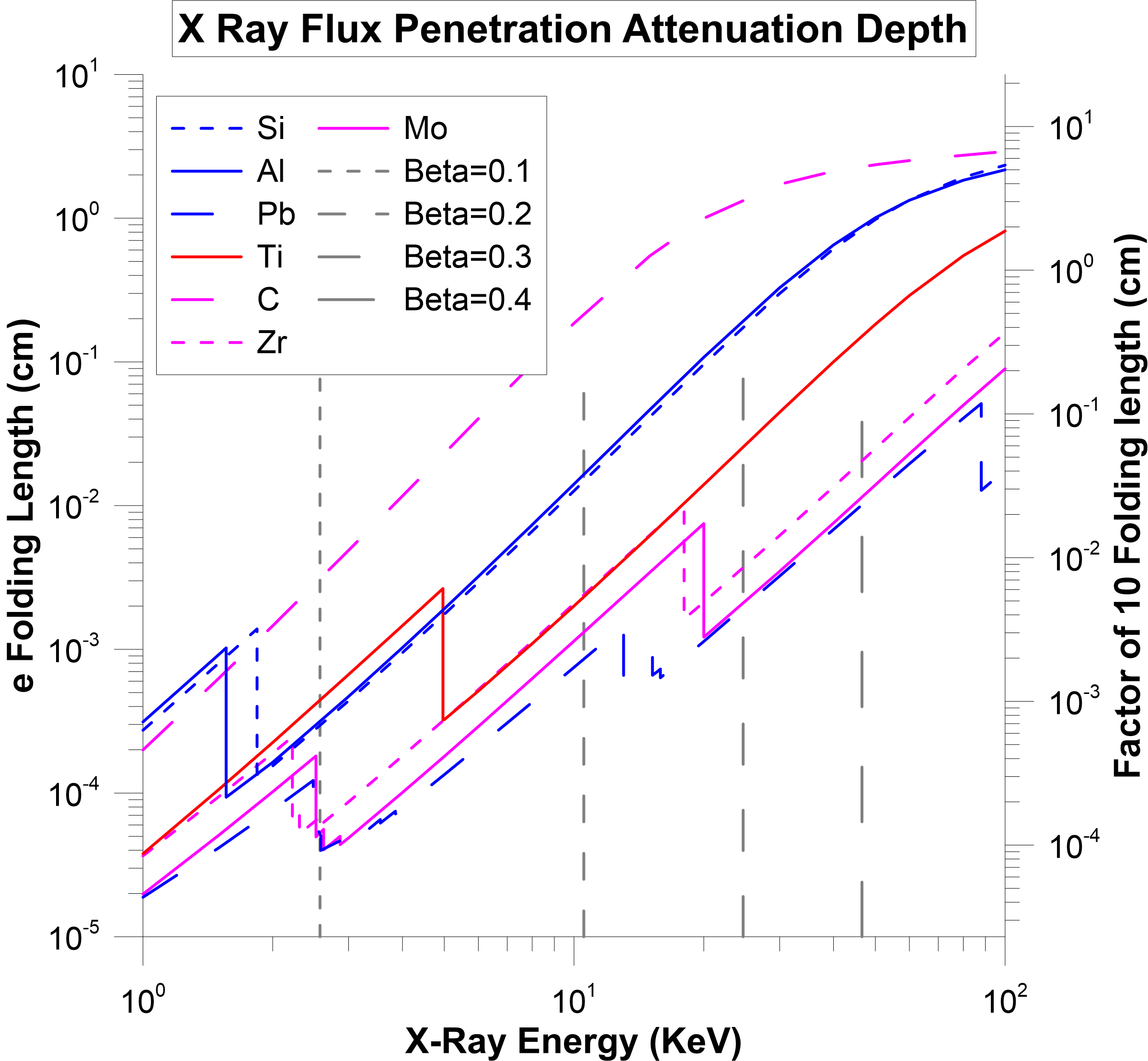} & \includegraphics[width=0.45\textwidth]{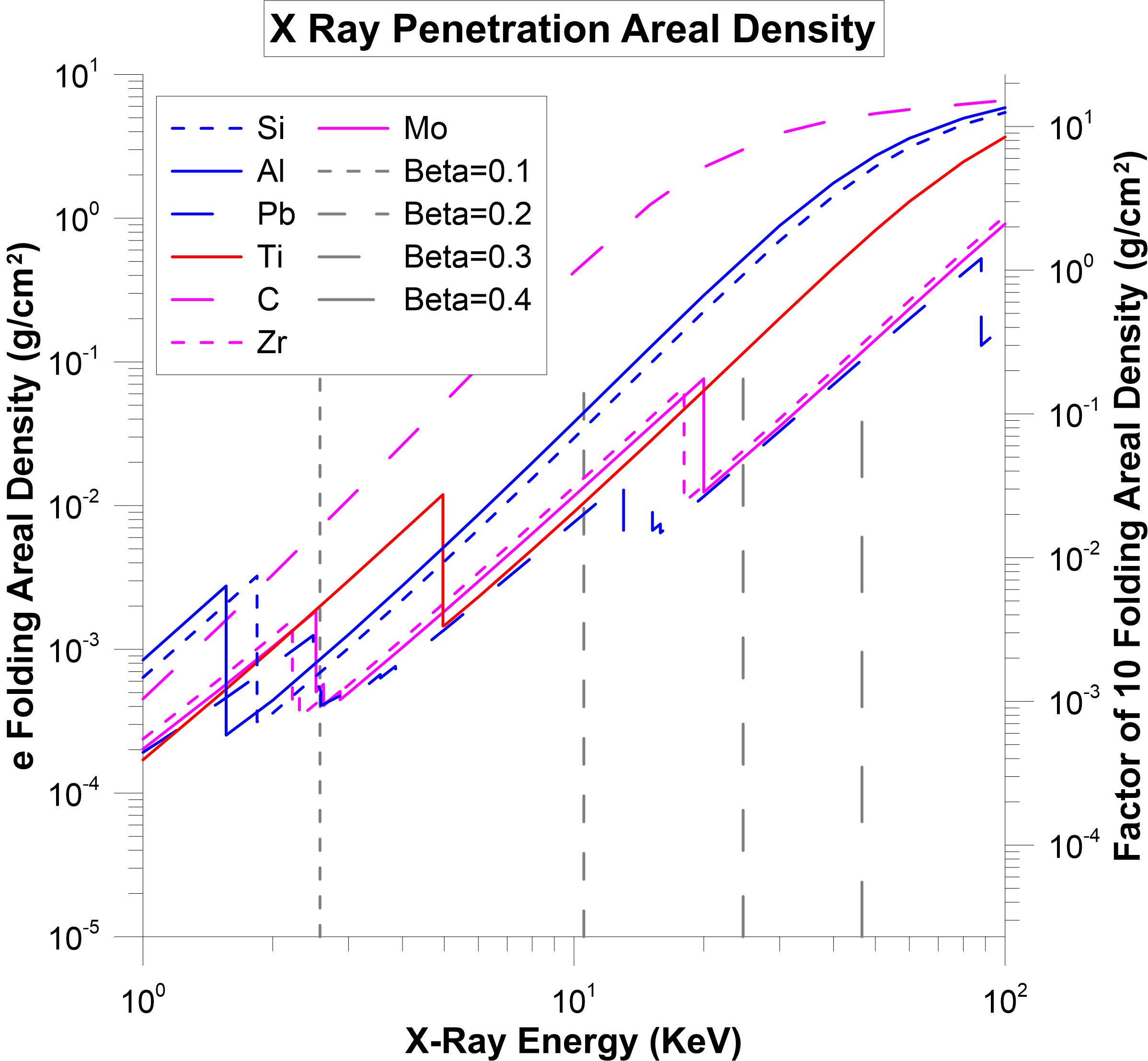} \\
             (a) & (b)
        \end{tabular}
        \caption{X-ray flux penetration length (a) and areal density (b) as functions of x-ray energy in various materials are given. The $\beta$ given is a worst case assuming all the energy from an incident ISM electron goes into a single photon. For incident protons, the worst case equivalent is assuming a ``head on'' collision of the  proton with an electron in the material and that all of the (hit) electron's energy goes into a single photon. In this case, the electron has twice the speed and four times the energy in the non-relativistic limit.}
        \label{fig:x-ray_penetration_and_areal_density}
    \end{figure*}

    We can describe the attenuation of an assumed idealized x-ray beam as being exponentially attenuated in both flux and in energy as:
    \begin{equation}
        I(z)=I(0)e^{-z/\alpha}
    \end{equation}
    where $I(z)$ is the flux at penetration depth $z$, and $\alpha$ is the flux attenuation at $e$ folding length. We can rewrite this as
    \begin{equation}
        I(z)=I(0)10^{-z/\alpha_{10}}
    \end{equation}
    where $\alpha_{10}=\ln(10)\alpha\sim 2.303\alpha$ is the flux attenuation factor at 10 folding lengths. Therefore, the x-ray energy at penetration distance $z$ is given by
    \begin{equation}
        E(z)=E(0)e^{-z/\alpha_{\textrm{energy}}}
    \end{equation}
    where $\alpha_{\textrm{energy}}$ is the energy attenuation at $e$ folding lengths (See Figure \ref{fig:x-ray_penetration_and_areal_density}).
    
    The two basic interactions are incident electrons on material electrons and incident protons on material electrons. As discussed above, electron penetration is extremely shallow (typ $<$ 1 micron) while proton penetration is significantly larger (typ $<$ 1 mm) at the spacecraft speeds we are exploring ($\beta<0.5$). Since the primary x-ray production comes from the $e_\textrm{inc}$--$e_\textrm{mat}$ or $p_\textrm{inc}$--$e_\textrm{mat}$ interactions, from both $e_\textrm{mat}$ bremsstrahlung and $e_\textrm{mat}$ inner shell transition with little direct $p_{inc}$ bremsstrahlung as the primary loss mechanism in both incident $e$ and $p$ is with the material electrons and not with the material nuclei as shown above.
    
    The x-ray production is largely isotropic and produced in a small region at the front leading edge. Since the nominal flight attitude strategy is to fly edge on this means that the number of x-rays incident on the spacecraft electronics is greatly reduced due to view factor (geometry) effects. We calculate the x-ray attenuation in the material below and show the overall x-rays the actually can impact the electronics in the spacecraft is very low.
    
    The maximum x-ray energy produced by incident electrons is the energy of the incident electron itself as the incident electrons can rapidly transfer their energy to the material electrons with the bremsstrahlung production spectrum falling rapidly for x-ray energies approaching the incident electron energy. The worst case (maximum penetration) is for the x-ray energy being equal to the incident electron energy (zero probability) and thus we assess this worst case below.
    
\subsection{Proton Bremsstrahlung}

    The situation with incident protons is different than that of incident electrons since the proton mass is so much larger than the material electrons it interacts with (p-nucleus interactions are greatly suppressed). The maximum material electron speed is very close to twice the incident speed of the incident proton though the distribution function is highly impact parameter dependent and thus largely weighted at lower ``electron ejection'' speeds. The worst case would be to assume interaction electrons come off at twice the speed of the incident proton (spacecraft speed). In general this is extremely improbable with the typical interaction electron energy being much less than 1 keV and thus x-ray energy production being greatly suppressed.
    
\begin{figure}[h]
        \centering
        \includegraphics[width=0.44\textwidth]{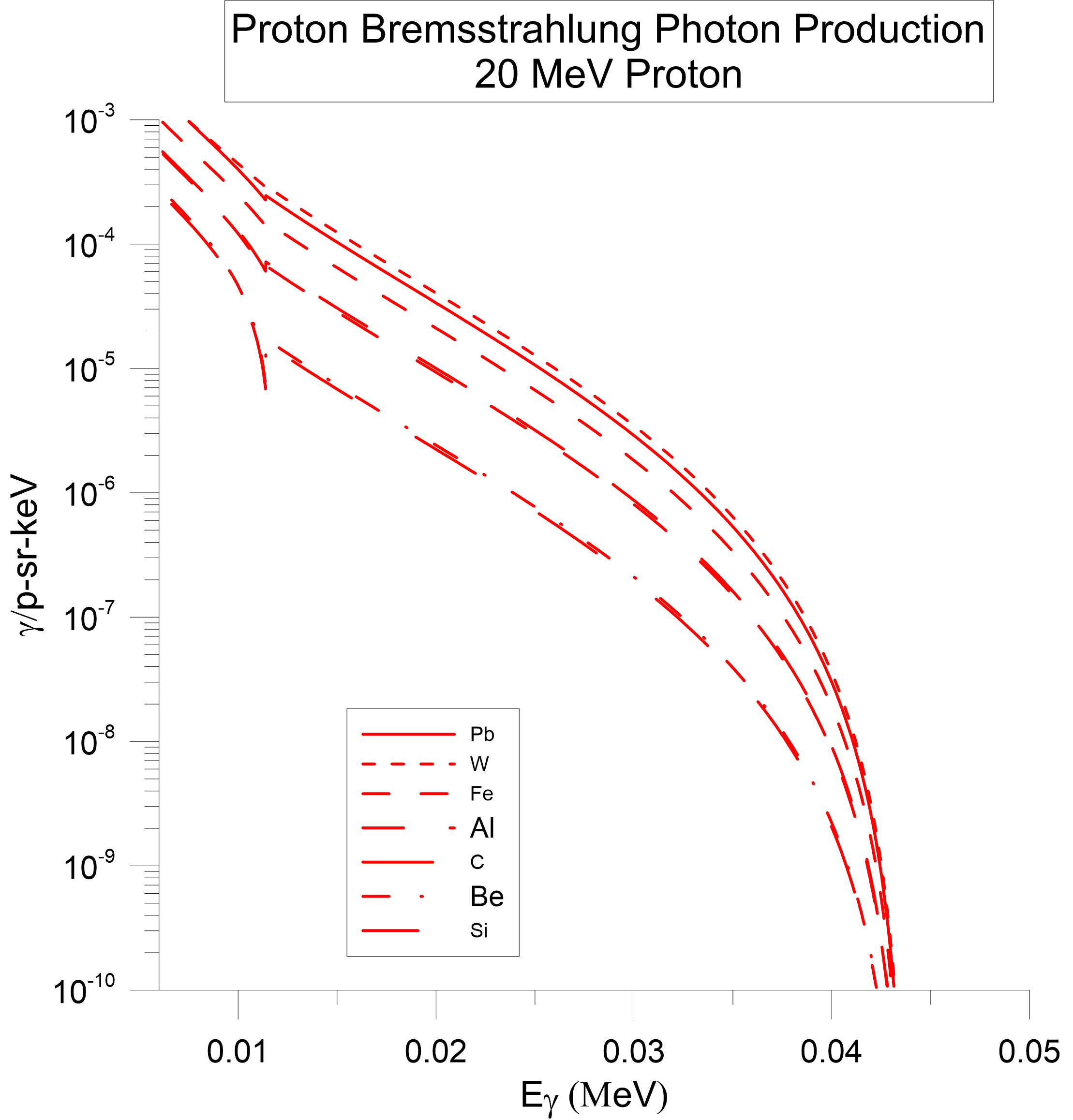}
        \caption{Bremsstrahlung photon production in units of photons per proton, per steradian, per keV for 20 MeV protons incident on a range of materials. A quasi-discontinuity can be seen in the vicinity of 0.01 MeV, where the two photon production mechanisms overlap. For photon energies below $\sim0.01$ MeV, the dominant production mechanism is the electron acceleration due to the initial proton-electron collision. For photon energies above $\sim0.01$ MeV, stopping of the accelerated electrons in the target material is the dominant bremsstrahlung production mechanism.}
        \label{fig:protonbremproduction}
    \end{figure}

    To determine x-ray production as a function of incident proton energy, we largely follow \citep{ogier_electron_1966}. Unlike for electrons, the energy loss of the proton due to bremsstrahlung is insignificant compared to collision-induced energy loss. Thus, for electrons and protons at equal $\beta$, although the protons have 2000$\times$ more energy than the electrons, few x-rays are produced due to proton bremsstrahlung. However, the proton loses energy predominantly through collisions with valence shell electrons in the material (not through nucleic interactions). Hence, the bremsstrahlung from these electrons \textit{is} a significant source of x-rays. This bremsstrahlung can be split into two separate effects. The first is the initial acceleration due to the proton-electron collision. Here, we use an experimentally confirmed production cross-section derived by Kramers for electron-nucleus collisions where the nucleus is stationary \citep{kramers_xciii_1923}. Therefore, the natural frame to use is the proton frame. Although in the material frame the electron's maximum speed is nearly twice the proton's (and hence the maximum energy is nearly four times that of the proton), in the proton frame this electron's maximum speed will be nearly equal to the proton's. Thus, the maximum energy---assuming a completely elastic collision, where the proton's energy change is negligible for a single collision---that a produced photon could have is given by $h\nu=mv^2/2$, where $v$ is the colliding proton's speed. The second portion of bremsstrahlung radiation production is due to stopping of the electrons in the material (i.e. deceleration of the electron). Now, the natural frame to use is the material frame, in which the electron maximally has twice the speed of the colliding proton. The maximum energy of an emitted photon is then four times that in the previous case. Each of these contributions has different cross-sections and, therefore, different integrals determining the amount of photons produced. Neglecting directional effects, we obtain the following formula for the differential number of photons at given energy $E_\gamma$ and solid angle $\Omega$ produced by bremsstrahlung of electrons during acceleration (i.e. increase in electron energy) that the incident proton collided with,
    
    \begin{align}
    \begin{split}
        \dfrac{dN_\gamma}{dE_\gamma d\Omega}(E_\gamma) &= 1.58\frac{Z'}{4\pi AE_\gamma} \\
        &\times\int_{1837E_\gamma}^{E_{p_0}} \frac{e^{-m(E)\cdot\mu /\rho}}{s(E)E} dE,
    \end{split}
    \end{align}
    where $Z'$ is the number of electrons in the material available for scattering (assumed equal to $Z$ the atomic number), $A$ is the atomic mass number of the material, $s$ is the stopping power in units of $\si{\mega\electronvolt cm^{2}\per\gram}$ as a function of proton energy $E$, which was obtained from NIST for the various materials considered \citep{berger_stopping-power_2009}, $E_{p_0}$ is the initial energy of the proton when colliding with the material, and $\mu/\rho$ is the mass absorption coefficient where $m(E)$ is given by the following formula. Note that the number 1837 is the ratio of the proton rest mass to the electron mass.
        \begin{equation}
        m(E) = (\cos ({\theta _p})/\cos ({\theta _q}))\mathop \int_E^{{E_{{p_0}}}} \frac{1}{{s(E')}}dE',
    \end{equation}
    where $\theta_p$ is the angle between the incident proton vector and the surface normal upon impact and $\theta_q$ is the angle between the emitted x-ray and the surface normal. Since we are not interested in attenuation of x-rays in the material along the incident proton path (given by the exponential term), we have set $m\mu/\rho\equiv 0$. Thus, the differential number of photons at a given energy produced by bremsstrahlung of electrons during deceleration (i.e. decrease in electron energy) due to material interactions is given by
    \begin{align}
    \begin{split}
        \dfrac{dN_\gamma}{dE_\gamma d\Omega}(E_\gamma)&=0.132\dfrac{ZZ'}{4\pi AE_\gamma}\int_{459E_\gamma}^{E_{p_0}}\dfrac{1}{Es(E)}\\
        &\times\left[\ln\left(\dfrac{E}{459E_\gamma}\right)+\dfrac{459E_\gamma}{E}-1\right]dE,
    \end{split}
    \end{align}
    where the attenuation term has been set equal to one, as mentioned earlier. Both bremsstrahlung photon-producing effects are plotted in Figure \ref{fig:protonbremproduction}.

    We plot the x-ray flux attenuation lengths vs x-ray energy, for various relevant materials in Figure \ref{fig:x-ray_penetration_and_areal_density}. We also show the maximum x-ray energy which is the incident electron energy for various $\beta$. As discussed, the typical x-ray energy is significantly lower than this. Compared to the incident e and proton beam power levels we computed above we note that the x-ray power is greatly reduced by various factors including electron-phonon interaction (lattice heating) and then the actual device irradiation is vastly reduced by both view factors (geometry) and by material attenuation as seen in the plot. The ``edges'' or jumps in the x-ray attenuation are due to material inner shell edges (typically K shell). The leading edge can be optimized to reduce both x-ray production as well as to attenuate x-rays that are produced through an optimized layered material approach. For example we see the factor of 10 attenuation length for Pb, Mo and Ti at $\beta=0.2$ are all quite similar with a length scale of about 50$\mu$m while at $\beta=0.3$ Pb and Mo are similar (also both are high density) while Ti is not as effective per unit length, but Ti has lower density. We also show the analogous plots but for areal density. Here the least mass solution (if x-ray attenuation was the only parameter to optimize - it is not) is similar for Pb, Ti, Mo and Zr at $\beta=0.2$ but at $\beta=0.3$ we would choose Pb, Zr and Mo, but less so Ti. We would choose different materials for different speeds, but x-ray production, edge degradation due to radiation damage, etc. are also a critical parameters to optimize.  Testing of various material combinations for difference energy $e$ and $p$ beams to access actual x-ray production and device sensitivity will be needed, but to first order the x-ray issue does not appear to be significant especially for edge on geometries. 
    
\subsection{Dust Grain Bremsstrahlung}
    
    An interesting discussion arises when considering the bremsstrahlung photon production from dust grain impacts. As a simplified analysis, consider the average interstellar dust grain to be 1 $\mu$m in diameter. Assuming a conservative 1 proton/{\AA} density within the grain, we can approximate that there are $\sim10^{12}$ protons/grain. Assuming a density of $10^{-15}$ grains/cm$^{-3}$, we see that there are approximately $1000\times$ more ISM protons than protons bound in dust grains per unit volume. However, the distribution of dust grain sizes is roughly a power law of the form $n(a)\propto a^{-3.5}$, where $a$ is the dust grain size, so there will be more protons bound in smaller dust grains (approximately equal dust mass per logarithmic interval of size) \citep{mathis_size_1977}. Since we cannot extend below 1{\AA} in grain size, and there are only 4 orders of magnitude from 1 $\mu$m to 1{\AA}, we see that there can only be an additional factor of 4 more protons bound in smaller dust grains than 1 $\mu$m.  Therefore, we see that photon production from proton impacts will dominate production from dust grain impacts simply due to the large difference in the number of protons in the ISM vs. those bound in dust grains. 
    
    However, it is also important to consider the possibility of collective effects. Assuming the nominal grain described previously, it is interesting to compare the wavelength of an expected bremsstrahlung photon to the dimensions of the grain. Since the kinetic energy of the grain ($\sim10^{19}$ eV at $\beta=0.2$) is much larger than the binding energy of the atoms in the grain ($\sim10^8$ eV for carbon), the grain impacts the target as an ionized and unbound collection of electrons and nuclei. The electrons will produce bremsstrahlung photons with maximum energy $\sim10$ keV at $\beta=0.2$, which have a wavelength of $\sim0.1$ nm. Similarly, the protons will produce bremsstrahlung photons mostly with energies between 1 and 10 keV. Bound nuclei lose energy in the material as $dE/dx\sim z^2$, and will thus stop within a shorter range inside the target due to the protons remaining intact in the nucleus during the bremsstrahlung process and thus acting as ``one charge." In the case of charges that are loosely bound (such as in a dust grain where the charges are bound due to weak electronic binding and not nuclear binding) this is no longer true. Thus in the dust grain the nuclei behave independently and not collectively (i.e. not firmly bound). Additionally, the wavelengths of bremsstrahlung photons produced is at least $10^4$ times smaller than the grain itself. 
    
\section{Particle Production from ISM Bombardment of the Forward Edge}
\label{section:particleproduction}

    As we have discussed, the forward edge of the spacecraft undergoes ISM bombardment. In addition to the radiation damage ($dE/dx$) that we have calculated there is also secondary (and tertiary) particle production from the bombardment. With the ISM consisting largely of electrons, protons, and helium there are a number of in situ particles produced from the impacts. These include:
    \begin{itemize}
        \item[-]Photons from bremsstrahlung, as discussed
        \item[-]Neutrons from proton and helium interactions with the spacecraft nuclei
        \item[-]Spallation products from the nuclear breakup of the spacecraft nuclei
        \item[-]Radioactive activation of the spacecraft
    \end{itemize}
    
    Due to the short range of the incident electrons, protons, and helium, it is a good approximation to assume that the production is essentially a surface phenomenon initially, though neutrons and photons can penetrate deeply into the spacecraft. We use the same coordinate system as for the radiation does calculations with the spacecraft velocity vector in the ISM frame being along the $z$-axis. Assuming the ISM particles are nearly at rest in the ISM frame we have the same particle velocity definitions as in equations (8), (9), (10), and (11). We assume spacecraft motion along the $z$-axis with $c\bm{\beta}=(0,0,c\beta)$. For a given ISM species density $n_s$ (\# volume$^{-1}$), the species flux $\bm{F_s'}$ (\# s$^{-1}$ area$^{-1}$) in the spacecraft frame is:
    \begin{equation}
        \bm{F_s'}=-n_s\bm{v'}=-n_s c\bm{\beta}-\bm{v}\sim-n_s c\bm{\beta},
    \end{equation}{}for $|c\bm{\beta}|>>|\bm{v}|$ (particle speed in ISM frame is low).
    
    For a particle of type $i$ being produced, the production per unit leading edge area is:
    \begin{equation}
        \frac{d\Gamma_i}{dA}=f_i(E,\Omega,...)\bm{F_s'}\cdot\bm{n},
    \end{equation}{}where $f_i(E,\Omega,...)$ is the production fraction which may depend on a number of parameters such as:
    \begin{itemize}
        \item Energy of particle being produced (there may be a spectrum)
        \item Per unit energy bandwidth
        \item Per solid angle
    \end{itemize}{}
    This depends on the type of particle being produced (photons, neutrons, spallation, etc.). The production of particle type $i$ is the integral over the area $\Sigma_f$ of the forward facing edges of the spacecraft where $\Sigma_f\rightarrow(\bm{F_s'}\cdot\bm{n}<0)$.
    \begin{align}
    \begin{split}
        \Gamma_i=\int_{\Sigma_f}\frac{d\Gamma_i}{dA}\mathop{dA}=f_i(E,\Omega,...)\int_{\Sigma_f}\bm{F_s'}\cdot\bm{n}\mathop{dA}
    \end{split}{}
    \end{align}{}
    Note that in order to compute the radiation dose from the produced particle at a specific location in the spacecraft we would have to compute $d\Gamma_i/dA (\bm{X'})$ as a function of the position on the leading edge, integrate over the forward edge area, and then calculate the flux of the produced radiation at the desired target point $\bm{X_0}$, which depends inversely on the square of the distance $|\bm{X'}-\bm{X_0}|^{-2}$ assuming there is no attenuation between the production and reception points. If there is attenuation, then we need to factor this in as well.
    
    For the case of isotropic production with no attenuation between production and target point $\bm{X_0}$, the flux at the target point is:
    \begin{align}
    \begin{split}
        \Gamma_i(\bm{X_0})&=f_i(E...)\int_{\Sigma_f}\frac{\bm{F_s'}\cdot\bm{n}}{4\pi|\bm{X'}-\bm{X_0}|^{-2}}\mathop{dA}\\
        &=-n_scf_i(E...)\int_{\Sigma_f}\frac{\bm{\beta}\cdot\bm{n}}{4\pi|\bm{X'}-\bm{X_0}|^{-2}}\mathop{dA}.
    \end{split}{}
    \end{align}{}
    
\subsubsection{Example: ISM Electron Bremsstrahlung}

    Assume $n_e=1$ cm$^{-3}$ and $\beta=1/3$. We have:
    \begin{equation}
        \bm{F_s'}=-n_sc\bm{\beta}=-10^{10}\hspace{1mm}\bm{\hat{z}}/\textrm{cm$^2$-s}
    \end{equation}{}
    The photon production fraction is thus:
    \begin{equation}
        f_i(E,\Omega...)\equiv\frac{dN_{\gamma/e}}{\mathop{dE_\gamma}\mathop{d\Omega}}=\kappa Z^n\bigg(\frac{E_0}{E_\gamma}-1\bigg)^x.
    \end{equation}{}
    Therefore, the photon flux at any point within the spacecraft is given by:
    \begin{align}
    \begin{split}
        \Gamma_i&=f_i(E,\Omega...)\int_{\Sigma_f}\bm{F_s'}\cdot\bm{n}\mathop{dA}\\
        &=\frac{dN_{\gamma/e}}{\mathop{dE_\gamma}\mathop{d\Omega}}\int_{\Sigma_f}\bm{F_s'}\cdot\bm{n}\mathop{dA}\\
        &=\kappa Z^n\bigg(\frac{E_0}{E_\gamma}-1\bigg)^x\int_{\Sigma_f}\bm{F_s'}\cdot\bm{n}\mathop{dA}.
    \end{split}{}
    \end{align}{}
    Note that $\Gamma_i$ in this case is the number of bremsstrahlung photons per unit energy and unit solid angle at a given photon energy.
    
    Production of secondary particles inside the spacecraft shield can cause various deleterious effects, including exposing critical spacecraft components to higher levels of radiation than levels expected for a non-relativistic spacecraft, as in the cases of bremsstrahlung photon production and neutron production. Additionally, secondary production of massive charged particles such as protons and alpha particles can contribute to the issue of hydrogen implantation, which can lead to destructive morphological changes to the shield and spacecraft due to bubble formation, migration, and bursting, as has been discussed extensively in earlier work\citep{drobny_survivability_2020,drobny_damage_2021}.


\section{Cosmic Rays}
\label{section:cosmicrays}

\subsection{Cosmic Ray Bombardment}

    Cosmic rays are charged particles that are from both galactic (GCR) and extragalactic (EGCR) sources that extend over a vast range of energies. Though there are charged particles due to our Sun (solar wind) the term ``cosmic rays'' is normally used when referring to the galactic and extragalactic portion. The density of charged particles in our solar system is highly variable in both save and time due to solar activity and planetary magnetic traps, such as the Van Allen belts around the Earth.
    
    The distribution of cosmic rays is largely isotropic. Irradiation of the spacecraft electronics can lead to degradation and failure and hence must be considered. It is important to understand the radiation dose to achieve a high degree of confidence for long term missions. Relevant here is that in the transition between the solar system and the ISM that tends to lower the flux of lower energy ($<$ 1 GeV) protons. Since we spend relatively little time in the solar system for relativistic interstellar missions we concentrate on the cosmic rays and the ISM. 
    
\begin{figure*}
         \centering
        \includegraphics[width=0.9\textwidth]{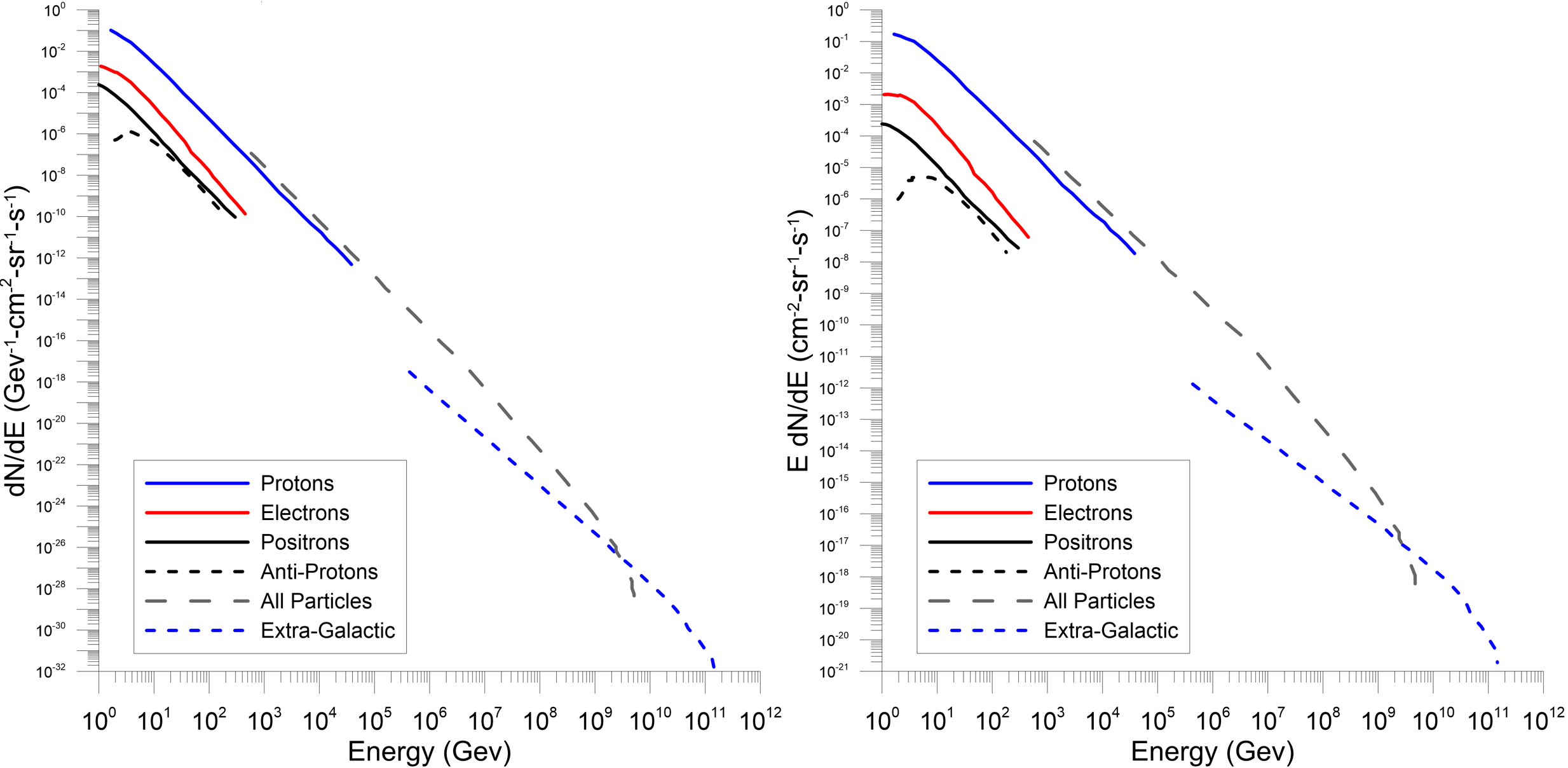} 
         \caption{Cosmic ray fluxes, $dN/dE$, for various cosmic ray species including the extra-galactic component (left). We also plot $E\cdot dN/dE$ (right) since for equal values of $E\cdot dN/dE$ the integral over $\log(E)$ intervals gives equal numbers of particles. Data obtained from \citep{maurin_database_2014, maurin_cosmic-ray_2020}.}
         \label{fig:cosmicrayfluxes}
     \end{figure*}

    We plot $dN/dE$ and $E\cdot dN/dE$ vs energy for the galactic component of cosmic ray protons, electrons, positrons, anti-protons as well as the extra galactic component in Figure \ref{fig:cosmicrayfluxes}, for which data was obtained from \citep{maurin_database_2014, maurin_cosmic-ray_2020}. We also plot $E\cdot dN/dE$ since for equal values of $E\cdot dN/dE$ the integral over $\log(E)$ intervals gives equal numbers of particles:
    \begin{equation}
        N(\textrm{$E_1\hspace{-1mm}\rightarrow\hspace{-1mm}E_2$})=\int_{E_1}^{E_2}E\frac{dN}{dE}\mathop{d(\ln{E})}=\int_{E_1}^{E_2}\frac{dN}{dE}\mathop{dE}.
    \end{equation}
    
    The flux of galactic comsic ray protons is well approximated by a power law from several GeV to beyond 100 TeV as: 
    \begin{equation}
        \frac{dN_z}{dE} (\textrm{cm}^{-2}\textrm{s}^{-1}\textrm{sr}^{-1}\textrm{GeV}^{-1})=a_z E(\textrm{GeV})^{-\alpha},
    \end{equation}
     where $a_1$=1.8 $(\textrm{cm}^{-2}\textrm{s}^{-1}\textrm{sr}^{-1}\textrm{GeV}^{-1})$ for protons and where $\alpha=\gamma +1=2.7$ is the differential spectral index and $\gamma=1.7$ is the integral spectral index. The spectral index $\alpha$ is largely independent of $z$ while $a_z$ is essentially just proportional to the fractional composition of the species ($z$) relative to $z=1$.  At E=1 GeV/nucleon  the power law above overestimates the flux by a factor of a few. Below several GeV per nucleon of kinetic energy the spectrum flattens to being nearly constant at 1 GeV. The Earth's local environment (magnetic field and solar effects) influences the lower energy spectrum so that local measurements show a distinct drop in flux below 1 GeV. As the cosmic ray flux ($dN/dE$) is dropping with increasing CR energy and since the energy loss per unit length ($dE/dx$) drops with increasing CR speed (CR energy) with a minimum $dE/dx$ at about 3 GeV per nucleon with a slow increase above that (``relativistic rise''), the primary energy deposition from cosmic rays will come from the lower energy portion of the spectrum (around 1 GeV).
    
     We can calculate the total flux $\Gamma$ (\# area$^{-1}$ s$^{-1}$) passing through one side of a surface by integrating over the projected area and over all energy from a lower energy cutoff $E_0$:
     
     \begin{equation}
         \Gamma(E_0 \rightarrow \infty)=\int_{2\pi} \int_{E_0}^{\infty}\cos(\theta)\frac{dN}{dE} \mathop{dE} \mathop{d\Omega}. 
     \end{equation}
     \\
     To compute the total energy deposition per unit length $d\xi/dx$ we would sum over all the CR species and integrate over the spectrum, solid angle and time.
     
     \begin{equation}
         \frac{d\xi}{dx}=\sum_z \int_E \frac{dN(z)}{dE}\frac{dE(z,E)}{dx}\cos(\theta) \mathop{dE} \mathop{d\Omega} \mathop{dt}.
     \end{equation}
     Here, the $\cos(\theta)$ term is the projected area where $\theta$ is the angle relative to the surface normal. Note that for relatively thin slices (thickness) and high energy, $dE/dx$ does not change much along the path. Hence for a ``slab or wafer'' of thickness $h$ the actual path length through the material at angle $\theta$ is $h/\cos(\theta)$. In this case we can calculate the energy deposited in thickness $h$ as:
     
     \begin{widetext}{}
     \begin{equation}
        \Delta E_{z,h}=\sum_z\int_E \frac{dN(z)}{dE}\frac{dE(z,E)}{dx}\frac{h}{\cos(\theta)}\cos(\theta) \mathop{dE} \mathop{d\Omega} \mathop{dt} = h\sum_z \int_E \frac{dN(z)}{dE}\frac{dE(z,E)}{dx} \mathop{dE} \mathop{d\Omega} \mathop{dt}.
     \end{equation}
     
     We calculate the exposure rate (for example Rad yr$^{-1}$) as:
     \begin{equation}
         R_{z,h}\bigg(\frac{d^{2}\xi}{\mathop{dx}\mathop{dt}}\bigg)=h\sum_z \int_E \frac{dN(z)}{dE}\frac{dE(z,E)}{dx} \mathop{dE} \mathop{d\Omega}.
     \end{equation}
     
     Assuming a power spectrum for cosmic ray species (a reasonable assumption based on available data for protons and higher $z$) for a given $z$ as $dN_z/dE=a_z E^{-\alpha}$, where $\alpha$ is roughly independent of $z$, but that the normalization $a_z$ does not depend on $z$.
     
     Note that when integrating over the forward hemisphere:
     \begin{equation}
         \int_{2\pi} d\Omega = \int\int\sin(\theta) \mathop{d\theta} \mathop{d\phi} = 2\pi, \hspace{3mm} \int_{2\pi} \cos(\theta) \mathop{d\Omega}=\pi, \hspace{2mm} \textrm{and}\hspace{2mm}\int_{2\pi}\cos^{2}(\theta) \mathop{d\Omega}=\frac{2\pi}{3}.
     \end{equation}
     \begin{align}
     \begin{split}    
         \Gamma_z(E_0 \rightarrow \infty)=\int_{2\pi}\int_{E_0}^{\infty}\cos(\theta)\frac{dN}{dE}\mathop{dE}\mathop{d\Omega}=\int_{2\pi}\cos(\theta)\mathop{d\Omega}\int_{E_0}^{\infty}a_z E^{-\alpha} \mathop{dE} = \frac{\pi a_z}{-\alpha+1}E_0^{-\alpha}.
    \end{split}
     \end{align}
     
     Below we discuss the energy loss for charged particles and see the form for $dE/dx$ at high energies for protons and higher $z$ reaches a minimum around 3 GeV per nucleon (kinetic energy) and then has a slow relativistic rise above that. We will thus assume that $dE/dx$ is roughly constant and equal to $\delta_{0,m}(z)$ where $m$ refers to the material and $z$ refers to the incident particle (i.e. $z=1$ for protons).
     
    \begin{align}
    \begin{split}
        \Delta E_{z,h}(\tau)=h\sum_z\int_{E,2\pi}\frac{dN(z)}{dE}\frac{dE(z,E)}{dx} \mathop{dE} \mathop{d\Omega} \mathop{dt} &=h\sum_z \delta_{0,m}(z)\int_{E,2\pi}\frac{dN(z)}{dE} \mathop{dE} \mathop{d\Omega} \mathop{dt} \\&=2\pi h \tau \sum_z \delta_{0,m}(z)\frac{a_z}{-\alpha+1}E_0^{-\alpha}.
    \end{split}
    \end{align}
    \end{widetext}
    We calculate the energy deposition in the thickness $h$ or exposure rate (for example Rad yr$^{-1}$ when converted) as:\\
    \begin{equation}
        R_{z,h}\bigg(\frac{d^2 \xi}{dxdt}\bigg)=2\pi h \sum_z \delta_{0,m}(z)\frac{a_z}{-\alpha+1}E_0^{-\alpha}.
    \end{equation}
    As seen below, the GCR composition is dominated by protons, even when the scaling of $dE/dx\sim z^2$ is taken into account and thus computing the dose from protons alone allows a reasonable estimate from all $z$. If we will take the lower energy cutoff  $E_0$ = 1 Gev/nucleon we get $a_1\textrm{(protons)}=1.8$ and with $\alpha=2.7$ we have the energy deposited per unit time and area in the slab of thickness $h$:
    \begin{align}
    \begin{split}
        R_{1,h}&=2\pi h \delta_{0,m}(z=1)\frac{a_z}{-\alpha+1}E_0^{-\alpha}\\
        &\sim 6.7h \delta_{0,m}(z=1),
    \end{split}
    \end{align}
    since $a_z$ is (typically) in units of $\#/(\textrm{cm}^2-\textrm{s}-\textrm{sr}-\textrm{GeV})$. If $\delta_{0,m}$ is in units of MeV g$^{-1}$ cm$^{-2}$, then $R_{1,h}$ is in units of MeV g$^{-1}$ cm$^{-3}$.

    As shown below, $\delta_{0,m}\sim2$ MeV g$^{-1}$ cm$^{-2}$ and is largely independent of the target material. This gives $R_{1,h}\sim13$ MeV s$^{-1}$ g$^{-1}$ cm$^{-3}$ $=2$ pW g$^{-1}$ cm$^{-3}$. Using 1 rad = 10 $\mu$J g$^{-1}$ we get $R_{1,h}=0.2$ $\mu$Rad s$^{-1}$ $\sim0.6$ Rad yr$^{-1}$. In conclusion, for a 30 year mission, this would be about 200 Rad which is a small dose for modern electronics (though significant for humans).
    
    \begin{figure}
        \centering
        \includegraphics[width=0.45\textwidth]{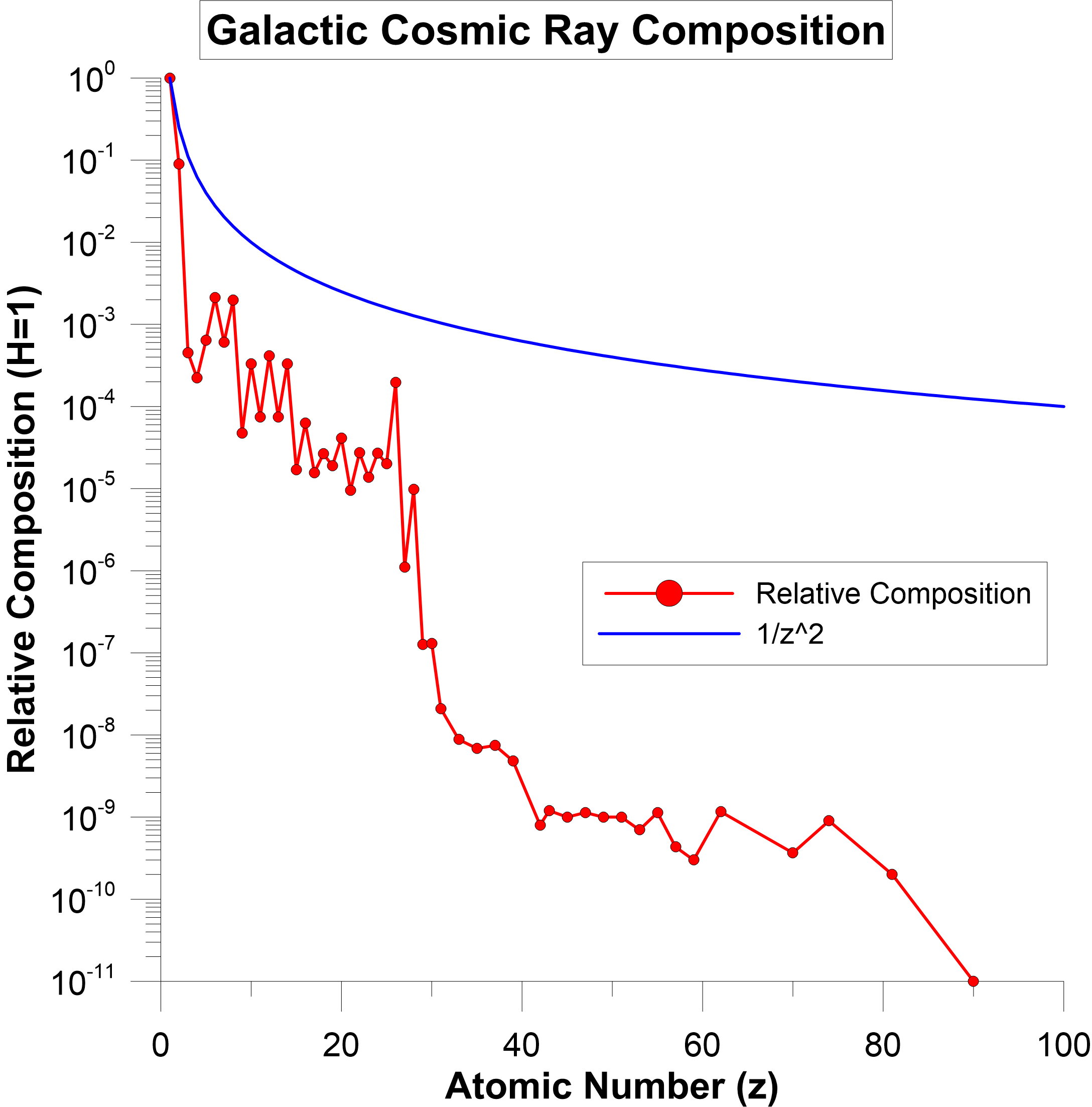}
        \caption{Galactic cosmic ray composition as a function of atomic number $z$ \citep{mewaldt_galactic_1994}. Also plotted in blue is the $1/z^2$ curve, which allows for a direct comparison to $dE/dx$, and it can be seen how the relative composition is always below the $1/z^2$ line, which implies that higher $z$ components of the cosmic ray spectrum do not significantly add to the total energy deposited in the spacecraft. This means that like the case for ISM, protons are the dominant source of energy deposition in the spacecraft from cosmic rays.}
        \label{fig:GalacticCosmicRayComposition}
    \end{figure}
    
    \begin{figure*}[]
    \centering
    \begin{tabular}{cc}
             \hspace{-10mm}\includegraphics[width=0.46\textwidth]{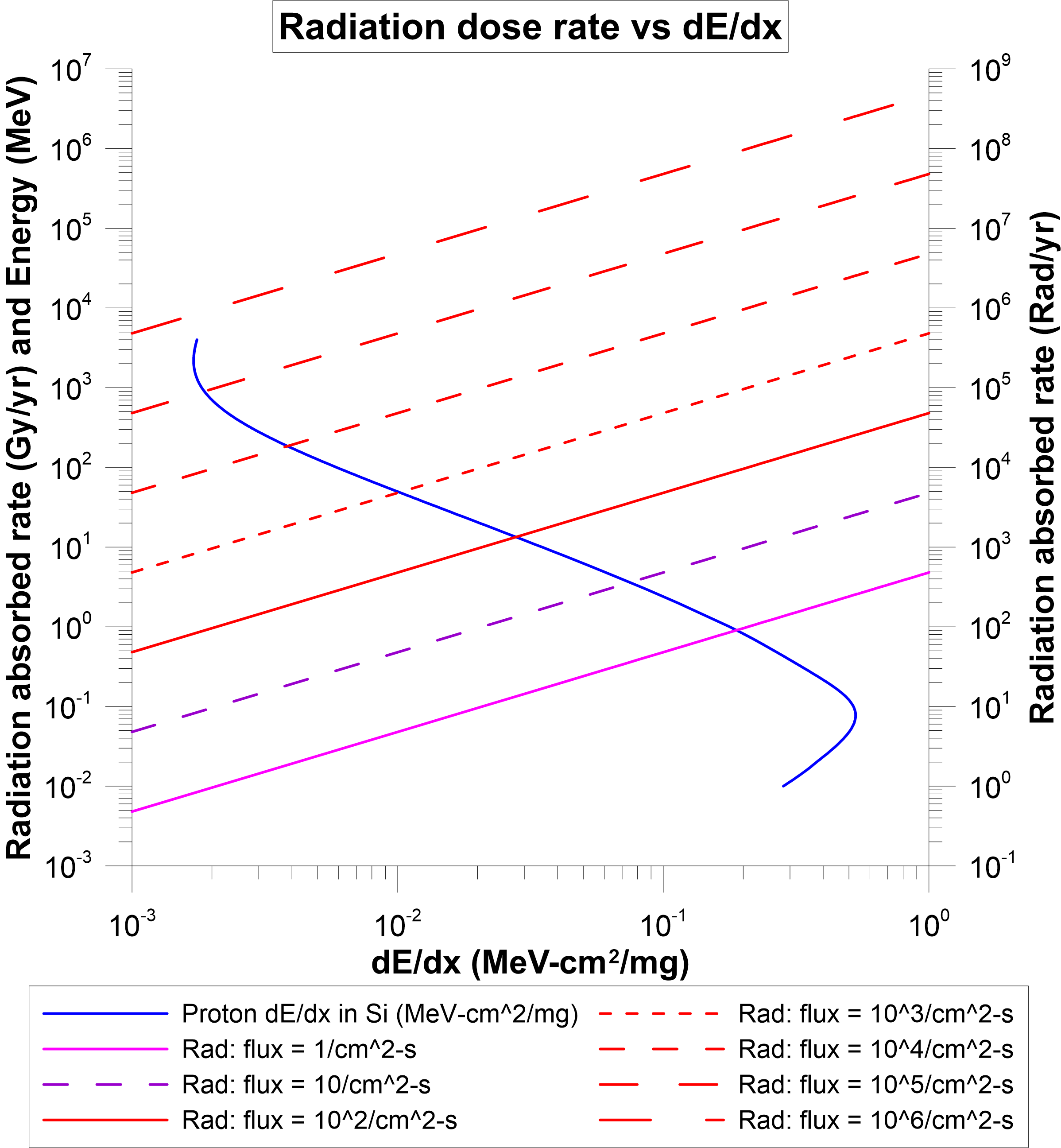} & \includegraphics[width=0.45\textwidth]{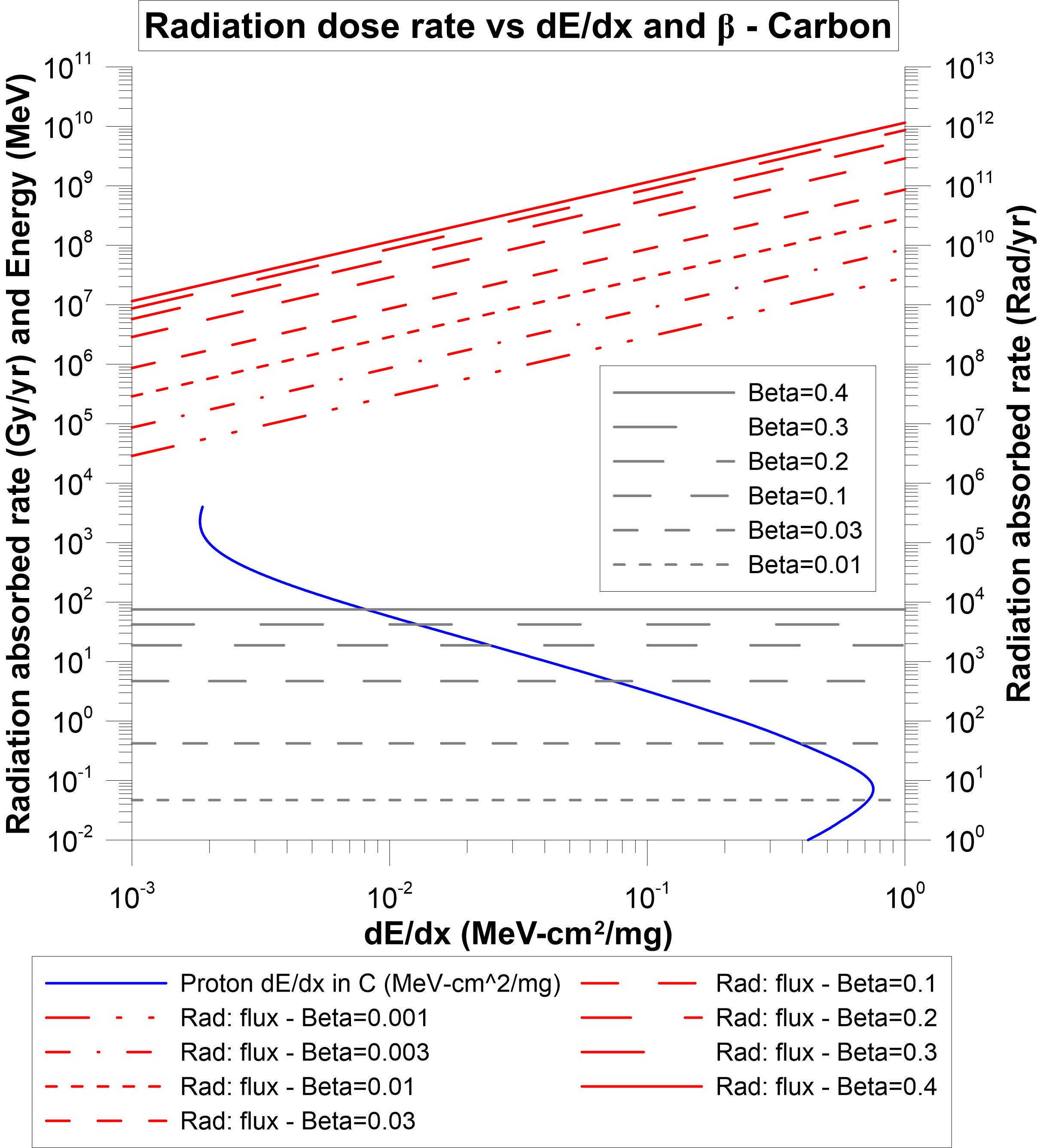} 
        \end{tabular}
    \caption{(Left) Radiation dose rate in silicon as a function of $dE/dx$ for proton radiation impact fluxes ranging from 1 to $10^6$ s$^{-1}$cm$^{-2}$, as described in the legend. (Right) Radiation absorption rate in carbon for varying spacecraft $\beta$ as a function of $dE/dx$ (red lines). Also shown as a blue curve in both plots is the proton stopping power $dE/dx$ in silicon (left) and carbon (right), which are to be read along the left vertical axis as $dE/dx$ as a function of proton energy in MeV. The grey lines represent incoming proton energies in MeV corresponding to spacecraft $\beta$ ranging from 0.01 to 0.4.}
    \label{fig:rad_vs_dEdx_imp-flux}
\end{figure*}
    
    Even in the frame of the spacecraft the galactic cosmic rays are nearly isotropic, unlike the boosted ISM which is highly peaked in the forward direction (below) and hence, unlike the case for the ISM, directional shielding for GCR's is not very effective, though shielding against GCR's is not normally needed as the dose is so low. The boosted ISM is a much larger problem.
     
\subsection{Cosmic Ray Composition}

    The composition of galactic cosmic rays is dominated by protons, but the other species are important to consider since the energy deposition $dE/dx\sim z^2/\beta^2$ and hence higher $z$ cosmic rays can be important if their flux is high enough. The spectra vs $z$ has been measured with reasonable accuracy and contains within the details of star formation and destruction as well as details about the acceleration mechanisms. About 79\% of the (number) of nucleons are in protons and 70\% of the remaining are in He with the rest in heavier elements. In Figure \ref{fig:GalacticCosmicRayComposition}, we plot the normalized composition of galactic cosmic rays vs $z$. We also show 1/$z^2$ to allow a comparison with $dE/dx$. 
    
    The fact that relative composition is below the 1/$z^2$ line means that higher $z$ components do not significantly add to the energy deposited in the spacecraft compared to protons for the same particle $\beta$. Typically the cosmic ray spectra for higher $z$ are given as energy per nucleon (total KE A$^{-1}$) and thus for the same $\beta$. The energy density of cosmic rays is about 1.4 eV cm$^{-3}$ which is curiously close to that of the CMB, starlight, the galactic and ISM magnetic field, radiation from dust, the thermal kinetic energy in the ISM, and even turbulence.

\section{Radiation Doses}
\label{section:radiationdoses}

\subsection{Device Radiation Tolerance}

    Normal radiation tolerance of commercial Si devices is typically about 10 Krad while special space SOI and SOS radiation hardened devices can withstand more than 1 Mrad with specialized devices for particle accelerators, such as the LHC at CERN, that can withstand more than 10 Mrad year$^{-1}$. This is somewhat larger than the doses we expect, but there is need to understand the long term multi-decade radiation resistance of the devices to be used. This is an area where much more research and development is possible with new semiconductors as well as an area that allows devices to be tested in a proton beam line prior to launch. In addition, the systems we envision all have multiple redundant systems as part of the baseline. Recent work on ``self-annealing'' devices using thermal annealing are also a promising area to explore if needed.

\subsection{Rads, Stopping Power, and Hit Flux}

    We can convert between the radiation absorbed $R$ (rad in cgs unit) or Gray (Gy in SI unit; 1 Gy = 100 rad = 1 J kg$^{-1}$) and the impact flux $\Gamma_p$ (\# s$^{-1}$m$^{-2}$) and $dE/dx$ for the impacting particle, energy, material, and exposure time $\tau$ with:
    \begin{align}
    \begin{split}
        R=\Gamma_p\frac{dE}{dx}\tau.
    \end{split}
    \end{align}
    This is shown in Figure \ref{fig:rad_vs_dEdx_imp-flux}, where we plot the radiation dose rate in silicon vs $dE/dx$ for various radiation impact fluxes, and the radiation absorption rate in carbon for varying $\beta$.

\subsection{ISM Impact Battery}

    As discussed in our ``roadmap'' \citep{lubin_roadmap_2016}, one interesting application is to convert some of the spacecraft kinetic energy into onboard power during the cruise phase via ISM proton and electron bombardment that heats the forward edge to form a thermal battery. This is the equivalent of an RTG but where the radioisotope heat source is replaced by ISM heating. Knowing the ISM density $n$ and the spacecraft speed $v$ we can proceed as above with the $dE/dx$ and range calculations to design an ISM Impact Battery.

\subsection{ISM Radiation Effects on Reflector}

    We can calculate the radiation dose on the reflector if it is deployed face on (normal vector and velocity vector parallel). Assuming an ISM proton density of $n=0.2$ protons cm$^{-3}$ we get $\Gamma_p = nv =0.2\beta c=6\times10^9 \beta$ (p s$^{-1}$ cm$^{-2}$). Over $\tau=1$ year this is a total of $\tau\Gamma_p = 2\times10^{17}\beta$ (p yr$^{-1}$ cm$^{-2})=20\beta$ (p yr$^{-1}$ \AA$^{-2}$). This is a formidable hit rate with twenty proton hits per \AA \hspace{1mm}cell. As was discussed earlier, for both incident protons and electrons, bremsstrahlung photon production occurs primarily due to interaction with material electrons and can produce cascades of x-rays, though typically more than 99\% of the electron bombardment energy  is  dissipated  as  heat  rather than radiation. However, considering that the atomic spacing is roughly an Angstrom, we see that a hit rate of $20\beta$ p-yr$^{-1}$-\AA$^{-2}$ is likely to cause non-negligible radiation production, as well as damage due to gas implantation \citep{drobny_survivability_2020}\citep{drobny_damage_2021}. If we used a graphene mono-layer for example we would have many hits per carbon atom. This is an area that would need to be measured for all reflector materials that would be used in a constant or long term deployed mode. This also applies to any structure, whether reflector or not, when oriented face-on.


\section{\label{sec:conclusion}Conclusion}

As has been shown, the radiation environment for a relativistic spacecraft is unique in relation to that experienced by the average Earth-orbiting or interplanetary spacecraft. While traversing the interstellar medium, a relativistic spacecraft will have to endure leading edge impacts with ISM particles while sustaining minimal damage to components critical to the operation of the spacecraft. A key component of the radiation environment is the radiation produced within the spacecraft upon impact and subsequent stopping of charged ISM particles like protons and electrons. As was discussed, at $\beta\sim0.2$ electrons will penetrate merely nanometers below the leading edge surface, while protons will penetrate to depths on the order of millimeters. Both electrons and protons will produce bremsstrahlung photons which will need to be attenuated by the shield material in order to protect critical spacecraft components. While the bremsstrahlung production from incident electrons in small (less than a medical x-ray), incident protons have the ability to produce cascades of bremsstrahlung photons deeper below the surface. Material choice for the shield will be key in mitigating these damage mechanisms. 

In addition to incident ISM species, the spacecraft will also have to weather much higher energy, though much less frequent, cosmic ray impacts. Even in the frame of a relativistic spacecraft, GeV/nucleon cosmic rays will impact practically isotropically, and thus a raised edge shield will do little to protect the spacecraft. However, due to their comparatively low flux, shielding schemes from cosmic rays for a relativistic spacecraft may not need to differ significantly from an ordinary spacecraft.  

\begin{acknowledgments}

Funding for this program comes from NASA grants NIAC Phase I DEEP-IN – 2015 NNX15AL91G and NASA NIAC Phase II DEIS – 2016 NNX16AL32G and the NASA California Space Grant NASA NNX10AT93H as well as a generous gift from the Emmett and Gladys W. fund.

\end{acknowledgments}

\bibliography{radeffects-references}{}
\bibliographystyle{aasjournal}
\nocite*



\end{document}